\documentclass[12pt]{article}
\usepackage{amsmath, amsfonts}
\usepackage{verbatim}   
\usepackage{epsfig,lscape}
\usepackage{float}
\usepackage{ natbib}
\usepackage{enumerate}
\usepackage{setspace}
\usepackage{graphics,graphicx}
\usepackage{array, graphicx}
\usepackage{epstopdf}

\usepackage{lscape}
\usepackage{color}
\usepackage{multirow}

\renewcommand{\baselinestretch}{1.33}

\newcommand{\RR}{{\bf R}}
\newcommand{\beq}{\begin{equation}}
\newcommand{\eeq}{\end{equation}}
\newcommand{\beqn}{\begin{eqnarray}}
\newcommand{\eeqn}{\end{eqnarray}}

\numberwithin{equation}{section}

\begin{document}
\renewcommand{\baselinestretch}{1.33}

\title{Bayesian Latent Variable Modeling of Longitudinal Family Data for Genetic Pleiotropy Studies}

\date{May, 2012}
\author{
Lizhen Xu  \thanks{Department of Statistics, University of Toronto, email: lizhen@utstat.toronto.edu}\\
Radu V. Craiu \thanks{Department of Statistics, University of Toronto,  email: craiu@utstat.toronto.edu} \\
Lei Sun \thanks{Division of Biostatistics, Dalla Lana School  of Public Health,  and  Department
of Statistics, University of Toronto, email: sun@utstat.toronto.edu}\\
Andrew D. Paterson \thanks{Program in Genetics and Genomic Biology, Hospital for Sick Children, and  Dalla Lana School  of Public Health, University of Toronto, Toronto, email: andrew.paterson@utoronto.ca}}

\maketitle
\pagebreak
\begin{abstract}
Motivated by genetic association studies of pleiotropy, we propose here a Bayesian latent variable approach to jointly study multiple outcomes or phenotypes. The proposed method models both continuous and binary phenotypes, and it accounts for serial and familial correlations when longitudinal and pedigree data have been collected. We present a Bayesian estimation method for the model parameters, and we develop a novel MCMC algorithm that builds upon  hierarchical centering and parameter expansion techniques to efficiently sample the posterior distribution. We discuss phenotype and model selection in the Bayesian setting, and we study the performance of two selection strategies  based on Bayes factors and spike-and-slab priors. We evaluate the proposed method via extensive simulations and demonstrate its utility with an application to a genome-wide association study of various complication phenotypes related to type 1 diabetes.

\end{abstract}


\section{Introduction} \label{introduction}

Pleiotropy occurs when a single genetic factor influences multiple continuous or binary phenotypes, and it is present in many genetic studies of complex human traits such as diabetes, hypertension and cardiovascular diseases. In genetic studies of complications or secondary manifestations of a disease, it is often believed that there are common genetic risk factors for the different phenotypes. In other cases, the primary and often conceptual phenotype (e.g. disease severity) may not be directly measured or be characterized by one single phenotype, and a set of surrogate response variables must instead be used. These response variables (phenotypes or outcomes) are mutually correlated as they measure the underlying trait from different perspectives. In order to take into account  all information and to increase statistical efficiency, it is desirable to model these outcomes jointly. 

An added characteristic of many emerging large-scale genetic studies is the collection of repeated measures over time in correlated individuals (family data). For example, the ongoing T2D-GENES (Type 2 Diabetes Genetic Exploration by Next-generation sequencing in multi-Ethnic Samples) consortium study
includes longitudinal measures of various T2D-revelant phenotypes (e.g.\ blood glucose levels and blood pressures) and other covariates (e.g.\ sex, body mass index and medication history), and these measures are collected for 2,500 individuals from 85 Mexican-American families. Similarly, the DCCT (Diabetes Control and Complications Trial) study includes longitudinal measures of various type 1 diabetes (T1D) complications, which we analyze in this paper. 

The longitudinal family studies combine the features of longitudinal studies in independent individuals and studies using single-time-point phenotype measures in families,  providing more information about the genetic and environmental factors associated with the traits of interest than cross-sectional studies \citep{Burton:2005ys}. However, joint modeling of multiple phenotypes using longitudinal family data involves non-trivial statistical and computational challenges because of the complex correlations that exist between different phenotypes (the phenotypical correlation), between repeated measures from the same phenotype (the serial correlation) and between individuals within the same family (the familial correlation).

Robust and powerful methods for the study of pleiotropy are under-developed in the literature due to data complexities that include a) the phenotypes of interest can be continuous, discrete or both, b) the joint effect of covariates on the multiple phenotypes is difficult to specify, and c) the familial, serial and other correlations are often present in the data as discussed above. There are a number of approaches proposed for cross-sectional data. For instance, \cite{Xu_etl:2003ys} extended the standard linear combination test to incorporate data-driven weighting factors, \cite{Welleretal:1996ys} applied the principal component analysis to the multiple traits of interest to obtain independent canonical variables and then conducted univariate quantitative trait locus (QTL)  analyses, and \cite{LangeWhittaker:2001ys} developed a QTL-mapping method based on generalized estimating equations. 

Here we propose to use the latent variable (LV) methodology to jointly study multiple correlated phenotypes in the presence of serial and familial correlations. The formulation of a latent variable model (LVM) relies on postulating the effect of a random variable that is not observed by the researchers but is assumed to exert an important influence on the set of  observed variables (also known as the manifest variables) and thus induces correlations among them  \citep{Bartholomew_2011_book}. In the context of pleiotropy studies, the manifest variables are the 
multiple observed phenotypes, which jointly inform the latent variable that represents the underlying conceptual disease status or severity. 

The LV methodology has been widely used in many scientific fields including economics, psychology and social sciences, and it is becoming increasingly attractive for genetic studies. For example, \cite{oHaraetal:2010ys} proposed a LV approach for the analysis of multivariate quantitative trait loci, \cite{Tayoetal:2008ys} applied a factor analysis (a sub-type of LVM) to find latent common genetic components of obesity traits, and \cite{Nocketal:2009ys} used the factor analysis for a metabolic syndrome study. 
Initial applications of LVM focused on reducing the number of manifest variables to a smaller number of latent outcomes.   \cite{SammelRyan:1996ys}  and \cite{SammelRyan:1997ys} extended the LVM to allow covariates to have effects on both the manifest and latent variables. \cite{RoyLin:2000ys} discussed a LV approach for longitudinal data with continuous outcomes. 

{
The proposed LVM consists of two parts. The first part models the relationship between the manifest variables and the LV to characterize the within-subject correlation among the different outcomes. The second part uses a linear mixed effect model to investigate the effect of a genetic marker and other covariates on the LV, accounting for the serial and familial correlations. Direct effects of covariates on the manifest variables are also allowed in the first part of model. 

The paper makes a number of  original contributions.  The proposed LV method generalizes the work of \cite{RoyLin:2000ys} to longitudinal family data with binary and continuous responses.  The Bayesian formulation can be desirable in practice as it  offers a principled way to incorporate prior information, often available in genetic studies, and to perform finite sample likelihood-based inference. However, the Bayesian model raises important computational challenges because  the sampling algorithm required to study the posterior is inefficient when it is applied in its standard form. We introduce a novel algorithm that relies on the hierarchical centering and parameter expansion techniques \citep{Gelfand:1995ys, LiuWu:1999ys, px-mvd} to improve computational efficiency.}

The rest of the paper is organized as follows. Section 2 introduces the details of the LVM and discusses the consequences of na\"{i}vely ignoring the family structure in the data. Section 3 presents a Bayesian estimation for the model parameters and  a novel MCMC algorithm designed to sample the posterior distribution efficiently.   Section 4 discusses phenotype selection and model selection in the Bayesian setting. Section 5 shows results from extensive simulation studies, and Section 6 applies the proposed method to a genetic study of T1D complications. Section 7 concludes with recommendations and further discussions.

\section{Latent Variable Model for Longitudinal Family data.} \label{section LVM model}

\subsection{The Statistical Model}

     Let $\textbf{Y}_{cit}=(y_{cit1}, \ldots, y_{citJ})^{'}$ be the $J\times 1$ vector of
  outcomes/phenotypes/manifest variables measured at the $t^{th}$ time on the $i^{th}$ individual
from the $c^{th}$ family/cluster for $c=1,2, \ldots, C$,
$i=1,2, \ldots, N_c$, $t=1,2, \ldots, T$, where C denotes
the total number of families, $N_c$ is the number of individuals
within the $c^{th}$ family, and $T$ is the total number of repeated
measurements. Among the $J$ outcomes, 
$\textbf{Y}^c_{cit}=(y^c_{cit1}, \ldots, y^c_{citJ_1})^{'}$ are continuous 
and $\textbf{Y}^b_{cit}=(y^b_{citJ_1+1}, \ldots, y^b_{citJ})^{'}$ are binary.
    Let $U_{cit}$ be the LV representing the conceptual disease
severity which aggregates the partial information brought by each of
the $J$ phenotypes. 

In the first part of the LVM, a continuous phenotype $y^c$ is linked to the
latent trait $U$ via a linear mixed model  
\begin{equation}\label{eqn:cts}
  y^c_{citj}=\beta_{0j}+W_{cit}^T\beta_{j}+\lambda_jU_{cit}+b_{cij}+e_{citj},
\end{equation}
where $e_{citj}\stackrel{\text{iid}}{\sim} N(0, \sigma^2_j) $,
$W_{cit}$ is a $p_1$-dimensional vector of covariates that have direct effects on the phenotype (direct fixed-effect covariates)
and  $\lambda_j$ is the factor loading that
represents the  effect of the LV on the $j^{th}$
phenotype. When all $\lambda_j$s are equal to 1,  model \eqref{eqn:cts} is reduced to a mixed effect model. The random component
$b_{cij}$ captures the family-specific within-subject correlations
over time. We assume $b_{cij}\stackrel{\text{iid}}{\sim} N(0,
\tau^2_j)$, and $e_{citj}$ and
$b_{cij}$ are mutually independent for $c=1, \ldots, C$, $i=1, \ldots, N_c$, $t=1, \ldots, T$ and $j=1, \ldots, J$. 

     If a phenotype is binary, a generalized linear mixed model is assumed, 
\begin{equation}\label{eqn:bin_linear_predictor}
 \eta_{citj}=\beta_{0j}+W_{cit}^T\beta_{j}+\lambda_jU_{cit}+b_{cij},
\end{equation}
with a probit link,
\begin{equation}\label{eqn:bin_link}
 E\left[y^b_{citj}|U_{cit},b_{cij}\right]=p(y^b_{citj}=1|U_{cit},b_{cij})=\Phi(\eta_{citj}).
\end{equation}
We choose a probit link instead of a logit link to gain
computational efficiency. Specifically, we take
advantage of the well-known representation of the probit model based
on a normally distributed LV. If
$\widetilde{y}^b_{citj}\sim N(\eta_{citj},1)$ is the  Gaussian variable underlying the
binary response $ y^b_{citj}$, then \eqref{eqn:bin_link} is recovered when
$y^b_{citj}=\mathbf{1}_{\{\widetilde{y}^b_{citj}>0\}}$.


The second part of the LVM assess the effect of $X_{cit}$, a $p_2-$dimensional vector of  variables that of primary interest (e.g.\ a genetic marker and possibly additional clinical factors), on the latent variable  $U$
via a linear mixed model,
\begin{equation}\label{eqn:latent}
U_{cit}=X_{cit}^T\alpha+Z_{cit}^T a_c+Q_{cit}^Td_{ci}+\epsilon_{cit},
\end{equation}
where $\epsilon_{cit}\stackrel{\text{iid}}{\sim} N(0,\psi^2)$.
Elements in $X$ are also called indirect fixed-effect covariates because their effects on a phenotype of interest $Y$ is induced via the effect of the latent variable  $U$ on $Y$ modelled in first part of the LVM above.
Pleiotropy is detected if the element of $\alpha$ corresponding to the genetic marker is found to be significant.
{
We assume $a_c\in \RR^{q_1\times 1}$ are the family-specific random effects and $Z_{cit}$ is the corresponding vector of covariates.  Similarly,  $d_{ci} \in \RR^{q_2\times 1}$ represents the subject-specific random effects with $Q_{cit}$ its associated covariate vector.  We assume that  $a_c\stackrel{\text{iid}}{\sim} MVN_{q1}(0, \Sigma_A)$,  $d_{ci}\stackrel{\text{iid}}{\sim} MVN_{q2}(0, \Sigma_D)$ and all random effects are independent of the  $\epsilon_{cit}$.}

\subsection{Model Identification Restrictions}\label{section model identification}
The following modification of
 equation (\ref{eqn:cts})
\begin{equation}\label{eqn:cts_modify}
  y^c_{citj}=\beta_{0j}+W_{cit}^T\beta_{j}+\lambda_jK^{-1}KU_{cit}+b_{cij}+e_{citj},
\end{equation}
where $K$ is an arbitrary nonzero number, suggests that, without any restriction on $\lambda$ or the variance of $U_{cit}$,  an infinite number of equivalent models can be created.
A similar phenomenon appears in the binary phenotype case. In 
order to avoid unidentifiability, we assume that the variance of $U_{cit}$ is equal to 1 and that  $\lambda_j$  is non-negative. 
Because we allow covariate effects in both parts of the LVM,  we assume that the two sets of covariates are disjoint and 
equation (\ref{eqn:latent}) does not contain the intercept.

\subsection{Effects of Ignoring Familial Correlation}

Individuals from the same family are genetically related resulting in correlation between their latent disease status. In practice, 
to reduce the analytic complexity and computation burden, one may choose to assume sample independence and apply existing methods (e.g.\ \cite{RoyLin:2000ys}).
However, ignoring the family structure will cause incorrect inference for the model parameters.
To see this clearly, we assume a simplified case where the phenotypes of interest are all continuous and there are no repeated measures. 
The two parts of LVM are reduced to 
$$y_{cij}=\beta_{0j}+W_{ci}^T\beta_{j}+\lambda_jU_{ci}+e_{cij}, \: \text{ and } \: \: \:
U_{ci}=X_{ci}^T\alpha+Z_{ci}^T a_c+\epsilon_{ci},$$
where $c=1, \ldots, C$, $i=1, \ldots, N_c$ and $j=1, \ldots, J$, with independent error terms
$e_{cij} \sim N(0,\sigma_j^2)$ and $\epsilon_{ci} \sim N(0,1)$,  $\lambda_j>0$ and $a_c \sim N(0,\Sigma_A$).

The variance of the $j^{th}$ response for individual $i$ from 
family $c$ can be 
decomposed in terms of the model parameters as:  
\begin{equation} \label{eqn:simplified_variance_withfamily}
Var(y_{cij})=\sigma^2_j+\lambda^2_jZ^T_{ci}\Sigma_AZ_{ci}+\lambda^2_j.
\end{equation}
Suppose we ignore the familial correlation in the data and propose the following model 
$$y_{hj}=\beta_{0j}+W_{h}^T\beta_{j}+\widetilde{\lambda}_j \widetilde{U}_{h}+e_{hj},
\: \text{ and } \: \: \:
\widetilde{U}_{h}=X_{h}^T \widetilde{\alpha}+\epsilon_{h},$$
where $h=1, \ldots, N$ is the individual's index and $N$ is the total sample size. 
In this case, the variance of the $j^{th}$ response for individual $h$ is decomposed as
 \begin{equation}\label{eqn:simplified_variance_nofamily}
Var(y_{hj})=\sigma^2_j+\widetilde{\lambda}^2_j.
\end{equation}
By equating 
\eqref{eqn:simplified_variance_withfamily} and \eqref{eqn:simplified_variance_nofamily}, 
 we find that $\widetilde{\lambda}^2_j=(Z^T_{ci}\Sigma_AZ_{ci}+1)\lambda^2_j$ and {{thus it is easy to see that}}
 $$\widetilde{\alpha}=\frac{\lambda_j}{\widetilde{\lambda_j}} \: \alpha= \frac{1}{\sqrt{(Z^T_{ci}\Sigma_AZ_{ci}+1)}}\: \alpha <\alpha.$$
Therefore, ignoring familial correlation can lead to 
significant underestimation of $\alpha$, the effect of a genetic marker 
on the LV of interest. However, the omission of existing familial correlation leads to overestimation of $\lambda$, the effect of the LV on the phenotypes of interest in the first part of the LVM, as demonstrated in the simulation studies of Section 5.2 below. This is consistent with what's reported in the statistical genetics literature (e.g.\ \cite{thornton}).
 
In the longitudinal setting,   as seen from equation (\ref{eqn:latent}), ignoring the familial correlation will also result in biased estimation of $\lambda$ and $\alpha$, as well as the serial correlation $\Sigma_D$. 
Simulation results reported in Section \ref{section pleiotropy simulation} support these conclusions. However, particular to  the longitudinal setting, if covariates $Z$ are not present,  
the error caused by ignoring the familial correlation can be absorbed into the serial correlation and thus only $\Sigma_D$ will be incorrectly estimated.
Nevertheless, there is still some
loss of efficiency in the estimation of $\lambda$ and $\alpha$ in this case.

\section{Parameter Estimation via Bayesian Method}\label{Bayes_est}

Traditional solutions for LVM, including the popular
software such as LISREL \citep{lisrel} and MPLUS \citep{mplus}, rely on frequentist methods.
 The development of modern computational algorithms, in particular Markov chain Monte Carlo (MCMC), 
enables us to use LVM for dependent data 
within the Bayesian paradigm. {
The  Bayesian approach also offers a principled approach to produce finite sample likelihood-based inference and to incorporate any available prior information which, in a genetic analysis setup, can be considerable. }

\subsection{Bayesian Model Setup}

The data in our model contain the observed continuous or binary outcomes ${\bf Y}$, the direct fixed-effect covariates $W$, the indirect fixed-effect covariates $X$, and the random-effect covariates $Z$ and $Q$. The vector of parameters of interest is $\Theta=({\beta_0}, {\boldsymbol \beta}, {\alpha}, {\lambda}, {\tau}^2, {\sigma}^2, \Sigma_A, \Sigma_D)^{'}$ where ${\beta_0}=(\beta_{01}, \ldots, \beta_{0J})^{'}$,
 ${\boldsymbol \beta}=(\beta_1^{'}, \ldots, \beta_J^{'})^{'}$ 
  where
 ${\beta_j}=(\beta_{j1}, \ldots, \beta_{jp_1})^{'}$,
 ${\alpha}=(\alpha_1, \ldots, \alpha_{p_2})^{'}$, 
${\lambda}=(\lambda_1, \ldots, \lambda_J)^{'}$,  ${\tau}^2=(\tau_1^2, \ldots, \tau_J^2)^{'}$ and ${\sigma}^2=(\sigma_1^2, \ldots, \sigma^2_{J_1})^{'}$. Therefore,
the posterior distribution for the model parameters is
$$p(\Theta|{\bf Y},W,X,Z,Q)\propto p(\Theta)p({\bf Y}|\Theta,W,X,Z,Q).$$

The complexity of the sampling model requires the use of MCMC algorithms for statistical inference. Unfortunately, the commonly used priors in probit and linear mixed effects models along with a run-of-the-mill sampling scheme lead to a torpidly mixing chain. In the next section, we discuss the  prior specifications  and the algorithmic modifications that we have implemented to improve the MCMC efficiency.

\subsection{MCMC sampling}
We follow  the data augmentation (DA) principle of \cite{tann-wong} and  sample alternatively from the posterior distribution given the complete data
and from the LV's distribution given the observed data and the parameter values. We first discuss the sampling scheme for the relatively simple case when all the phenotypes are continuous and we then extend the algorithm to binary phenotypes. 

\subsubsection{Continuous Traits}
When all the phenotypes are continuous and the conditional conjugate priors are defined for the model parameters, 
a 
 {standard Gibbs (SG) sampler} can be used for MCMC sampling from the posterior distribution. 
However, due to high dependence between the components of the Markov chain corresponding to the parameter vector $\mathbf{\Theta}$ and {the missing and latent data vector} $\Omega$,  we observe a very slow mixing of the chain. Improvement can be obtained by using hierarchical centering (HC)  \citep{Gelfand:1995ys}.  The HC technique moves the parameters up the hierarchy via model reformulation.
Specifically, in \eqref{eqn:cts} we move $\beta_{0j}$'s  up the model hierarchy to be the mean
of the random effect $b$ so that the new random effect 
is $b^*_{cij}=\beta_{0j}+b_{cij}$. 

Another technique devised to overcome the slow convergence problem of DA algorithms is parameter expansion (PX) \citep{px-mvd,LiuWu:1999ys}.
The idea behind PX is to introduce auxiliary parameters in the model and  average over all their possible values in order to produce inference for the original model of interest. As demonstrated by \cite{px-mvd} and \cite{LiuWu:1999ys}, the benefits of this apparently circuitous strategy can be highly significant, because  the larger parameter space allows the Markov chain to move more freely and  breaks the dependence between its components.
  
In our implementations, we have observed a strong coupling between the updates of the random effect $b_{cij}$ and its variance $\tau^2_j$, and between the factor loading $\lambda_j$ and the latent factor $U$ for all $1\le j \le J$.
 For instance, an update of $\tau^2_j$ close to zero likely yields a small update of $b_{cij}$ and vice versa. 
Thus, we introduce auxiliary parameters $\xi_j$ and $\psi$ and define the following {\it PX with hierarchical centering model (PX-HC)}
\begin{equation}\label{eqn:PX-HC-cts}
  y^c_{citj}=W_{cit}^T\beta_{j}+\lambda^*_jU^*_{cit}+\xi_jb^*_{cij}+e_{citj}, \mbox{ with}
\end{equation}
\begin{equation}\label{eqn:PX-HC-latent}
U^*_{cit}=X_{cit}^T\alpha^*+Z_{cit}^T a^*_c+Q_{cit}^Td^*_{ci}+\epsilon^*_{cit},
\end{equation}
where $b^*_{cij}\sim N(\beta^*_{0j},\eta^2_j)$, 
$a^*_c\stackrel{\text{iid}}{\sim} MVN_{q1}(0, \Sigma^*_a)$,
$d^*_{ci}\stackrel{\text{iid}}{\sim} MVN_{q2}(0, \Sigma^*_d)$, and $\epsilon^*_{cit}\stackrel{\text{iid}}{\sim} N(0,\psi^2)$. To relate the original parameters to the expanded parameters,
 the following transformations are used
$$ \alpha=\alpha^*/\psi,  \;\;\;\;\;  U_{cit}=U_{cit}^*/\psi,  \;\;\;\;\;  
 \Sigma_A=\Sigma^*_A/\psi^2,  \;\;\;\;\;  \Sigma_D=\Sigma^*_D/\psi^2,$$
$$\lambda_j=\lambda_j^*\psi,   \;\;\;\;\;  \beta_{j0}=\beta^*_{j0}\xi_j, \;\;\;\;\;  \tau^2_j=\xi^2_j\eta^{2}_j, \;\; \mbox{ for all } 1\le j \le J.$$

The conjugate priors used for the auxiliary parameters involved in the PX  scheme lead to a Gibbs sampler in which each conditional distribution is available in closed form and induce the  folded-t (the absolute value of $t$ distribution) prior distributions for the parameters $\tau$ and $\lambda$. More precisely, since  $\mbox{Var}(b^*_{cij})=\eta^2_j$ in our PX-HC model and $\mbox{Var}(b_{cij})=\tau^2_j$ in the original model, we have $\tau_j=\lvert {\xi_j} \rvert \eta_j$. When the conditional conjugate normal and inverse-Gamma prior are
applied to $\xi_j$ and $\eta^2_j$, respectively, the resulting prior for $\tau_j$ is the folded-t distribution.  Similarly, since
 $\lambda_j=\lambda_j^*\psi$, a  half normal prior assigned to  $\lambda_j^*$ and inverse-Gamma prior to $\psi^2$ will result in
a folded-t prior for $\lambda_j$. Other authors have 
discussed the suitability of folded-t priors in mixed effects and factor analysis models.
 For instance, \cite{Gelman:2006ys}
noted the added flexibility and improved behaviour when random effects are  small, and \cite{Ghosh:2009ys}  suggested  the use of $t$ or folded-t priors for the factor loadings in a factor analysis setting. 
%

We consider independent and conjugate priors for the expanded model parameters 
$\Theta^*=({\beta^*_0},{\boldsymbol \beta},{\alpha^*},{\lambda^*},{\eta}^2,{\sigma}^2,\Sigma^*_a,\Sigma^*_d,\psi,{\xi})^{'}$. 
 For example, we use normal priors for the fixed-effect coefficients. Thus,

\begin{description}
\item  $\beta^*_{0j} \sim N(0,1000),\; \beta_j \stackrel{\text{iid}}{\sim}N_{p_1}(\mathbf{0},1000*\mathbf{I}_{p_1}),\;  \eta^2_j \sim \mbox{Inv-Gamma}\left (\frac{v_2}{2},\frac{v_2}{2}\right)$, for $1\le j \le J$;
\item  $\alpha \sim N_{p_2}(\mathbf{0},1000*\mathbf{I}_{p_2})$.
\end{description}
For the scale parameters, we specify conditional conjugate priors,
\begin{description}
\item $\sigma_j^2 \sim \mbox{Inv-Gamma}(0.1,0.1)$ for $j=1,\ldots, J$;
\item $\Sigma^*_a \sim \mbox{Inv-Wishart}(V_a,S_a^{-1})$ with $V_a=q_1+1$ and $S_a=10*\mathbf{I}_{q1}$;
\item $\Sigma^*_d\sim \mbox{Inv-Wishart}(V_d,S_d^{-1})$ with $V_d=q_2+1$ and $S_d=10*\mathbf{I}_{q2}$.
\end{description}
For the auxiliary parameters we have introduced, the priors are 
\begin{description}
\item  $\psi^2 \sim \mbox{Inv-Gamma}\left (\frac{v_1}{2},\frac{v_1}{2}\right), \;\; \xi_j \sim  N(0,1), \;\; $
\end{description}
where $v_1$ and $v_2$ are the hyperparameters representing the degrees of freedom (df) of the induced folded-t priors for $\lambda_j$ and
$\tau^2_j$, respectively. 

For the purpose of phenotype selection, which will be discussed in detail in Section 4.1, we specify a spike-and-slab prior for $\lambda^*_j$ 
 \begin{equation}
 p(\lambda^*_j)= (1-\pi_j)\delta_{\{0\}}+\pi_j \mbox{TN}_{+}(0,1),
 \label{lambda-prior}
 \end{equation} 
 with the hyperparameter
$\pi_j \sim \text{Beta}(a,b).$ Notice that the induced prior for $\lambda$ in the original
inference model is also a spike-and-slab prior with the spike at zero and the slab distribution equal to a folded-t distribution with $v_1$ df. 
 
After defining the conditional conjugate priors for the expanded model, we now can apply the Gibbs sampling to obtain the posterior samples. Details of sampling are described in the Appendix, and the key steps at a given iteration $k-1$ are:
\begin{description}
 \item[ Step 1:] Draw $\Theta^{*(k)}$ from $f(\Theta^{*}|\Omega^{*(k-1)}, {\boldsymbol y^c})$, which involves sampling
 $(\beta^{'}_1,...,\beta^{'}_J)^{'}$, $\beta^*_0$, $\alpha^*$, $\lambda^*$, $\eta^2$, $\sigma^2$,
$\Sigma^*_a$, $\Sigma^*_d$, $\psi$, $\xi$, and
\item[ Step 2:] Draw $\Omega^{*(k)}$ from $f(\Omega^{*}|\Theta^{*(k)}, {\boldsymbol y^c})${, which involves sampling $\Omega^*=(U^*, b^*, a^*, d^*)^{'}$.}
\end{description}

\subsubsection{General Traits}
Suppose that the phenotypes of interest includes both continuous and binary traits. Without loss of generality, we assume that the first $J_1$ phenotypes are continuous and the remaining ones are binary. In order to address concerns involving the MCMC mixing similar to those in the continuous response 
case, we define 
the 
 model
\begin{equation}\label{eqn:PX-HC-cts_mixed}
  y^c_{citj}=W_{cit}^T\beta_{j}+\lambda^*_jU^*_{cit}+\xi_jb^*_{cij}+e_{citj}, \: \: 1\le j \le J_1,
\end{equation}
\begin{equation}\label{eqn:PX-HC-binary_mixed}
p(y^b_{citj}=1)=\Phi(W_{cit}^T\beta_{j}+\lambda^*_jU^*_{cit}+\xi_jb^*_{cij}), \: \: J_{1}+1\le j \le J,
\end{equation}
\begin{equation}\label{eqn:PX-HC-latent_mixed}
U^*_{cit}=X_{cit}^T\alpha^*+Z_{cit}^T a^*_c+Q_{cit}^Td^*_{ci}+\epsilon^*_{cit},
\end{equation}
where $b^*_{cij}\sim N(\beta^*_{0j},\eta^2_j)$, 
$a^*_c\stackrel{\text{iid}}{\sim} MVN_{q1}(0, \Sigma^*_a)$,
$d^*_{ci}\stackrel{\text{iid}}{\sim} MVN_{q2}(0, \Sigma^*_d)$, $\epsilon^*_{cit}\stackrel{\text{iid}}{\sim} N(0,\psi^2)$, and the prior distributions are the same as in the continuous case.

A specific issue encountered here is that some of the conditional distributions required in the Gibbs sampler are not available in closed form. 
One possible solution is to employ the DA scheme proposed by \cite{AlbertChib:1993ys} that uses the underlying continuous variable $\widetilde{y}^b_{citj}$ (discussed in Section 2.1) as an auxiliary variable, so that all the conditional posterior densities in the expanded model 
 can be directly sampled from. However, even in the model defined by \eqref{eqn:PX-HC-cts_mixed}$-$\eqref{eqn:PX-HC-latent_mixed} we have noticed  strong posterior dependence between $\widetilde{y}^{b}_{citj}$ and some of the model parameters which triggered a sluggish mixing of the chain. We apply an additional layer of the PX scheme by introducing a working parameter $\gamma_j$ and
a one-to-one mapping $\widetilde{y}^{b*}_{citj}=\gamma_j\widetilde{y}^b_{citj}$. 
We use the marginal DA scheme 3 of \cite{DykMeng:2001ys}, leaving the prior distributions and the parameterization
 of the model unchanged. We call this algorithm the {\it doubly parameter-expanded with hierarchical centering} and denote it as $\mbox{PX}^2-$HC.
  Below we summarize 
 the $k$th iteration in  the Gibbs sampling algorithm, and we provide a complete description in the Appendix.
\begin{description}
\item[ Step 1:] Draw 
\[
\widetilde{y}^{b(k)}_{citj}\sim \left \{
\begin{array}{cc}
TN_{+}(\mu^{*}_{citj},1),& \mbox{ if } y^b_{citj}=1\\
TN_{-}(\mu^{*}_{citj},1),& \mbox{ if } y^b_{citj}=0\\
\end{array}
\right .,
\]
where $\mu^{*}_{{citj}}=W_{cit}^T\beta^{(k-1)}_{j}+\lambda^{*(k-1)}_jU^{*(k-1)}_{cit}+\xi^{(k-1)}_jb^{*(k-1)}_{cij}$. Transform $\widetilde{y}^b_{citj}$ to $\widetilde{y}^{b*}_{citj}$  via $\widetilde{y}^{b*}_{citj}=\gamma_j\widetilde{y}^b_{citj}$.
\item[ Step 2:] For $j=J_1+1,...,J$, the order of updating $(\beta^{(k)}_j,\lambda^{*(k)}_j,\xi^{(k)}_j,\gamma^{2(k)}_j)$ involves sampling first $\gamma^{2(k)}_j$ and then $(\widetilde{\beta}^{(k)}_j,\widetilde{\lambda}^{*(k)}_j,\widetilde{\xi}^{(k)}_j)$ from their respective conditional densities.
Set $\beta^{(k)}_j=\widetilde{\beta}^{(k)}_j/\gamma^{(k)}_j$, $\lambda^{*(k)}_j=\widetilde{\lambda}^{*(k)}_j/\gamma^{(k)}_j$, and $\xi^{(k)}_j=\widetilde{\xi}^{(k)}_j/\gamma^{(k)}_j$.
       
The updates of the remaining parameters for the continuous responses and for the second part of the LVM are the same as those for the case with only continuous responses (see Section 3.2.1 and the Appendix).
\item[ Step 3:] Draw $\Omega^{*(k)}$ from $f(\Omega^{*}|\Theta^{*(k)},y^c, \widetilde{y}^{b*(k)}_{citj})$.
\end{description}
%
Figures 1 and 2 illustrates the significant improvement,  in
terms of reduced autocorrelation, 
brought by the additional layer of parameter expansion. 
See Section 5  for details of the simulations and the Appendix for  additional comparison plots.
\section{Model and Variable Selection}\label{model_selection}
\subsection{Selection of Relevant Phenotypes}

In medical research, it is of interest to determine if a phenotype under the study is truly relevant to the latent disease  status.  In the LVM setting, this is equivalent to testing if the coefficient or factor loading $\lambda_j$ for the $j^{th}$ phenotype (\eqref{eqn:cts} for continuous phenotype and \eqref{eqn:bin_linear_predictor} for binary phenotype)
is statistically significant or not. 


The sign restriction on the 
factor loadings, $\lambda_{j}s\ge 0$ as discussed in Section 2.2, implies that the highest posterior density interval (HpdI) will seldom include zero and is thus anti-conservative as a selection criterion.
To assess the significance of the factor loadings, we apply the Bayesian model selection method using the spike-and-slab priors \citep{Mitchell:1988ys, 
George:1993ys, Chipman:1996ys,XuCraiuSun:2011ys}. 
Specifically, we use the spike-and-slab prior \eqref{lambda-prior} for each  $\lambda_{j}^*$. 
 The point mass  at zero shrinks small values of the factor loading towards zero, while the values of $a$ and $b$ reflect the prior belief in $\lambda_j=0$ \citep[see][for a similar discussion on the use of spike-and-slab priors to alleviate the winner's curse in genetic association studies]{XuCraiuSun:2011ys}. When there is no prior information, we recommend $a=1$ and $b=1$ which correspond to $\pi_j \sim \text{Unif}(0,1)$.  For those phenotypes that are {\it a 
priori} considered to be unrelated to the latent disease variable, we can set $a=0.25$ and $b=1$, thus favouring {\it a priori} small values for the corresponding $\lambda_j$s.
 
 The determination of the relevance of the $j^{th}$ phenotype 
 is based on the posterior probability of positive loading, $Pr(\lambda_j>0|\boldsymbol Y)$. The sampling algorithm discussed in the Appendix introduces the latent mixture indicator $\omega_j=\mathbf{1}_{\{\lambda_j>0\}}$, so that
 $Pr(\lambda_j>0|\boldsymbol Y)$  can be approximated by the MCMC sample frequency of $\{\omega_j=1\}$. 
Given a pre-specified threshold $\phi$, we assume that any loading with $Pr(\lambda_j> 0|\boldsymbol Y)\ge\phi$ should be included in the model. The value for  $\phi$ depends on the practical problem. 
If the number of manifest variables is large, the $\phi$ value can be chosen to control the overall average Bayesian false discovery rate (FDR) \citep{Morris:2008ys}. The performance of selecting the correct phenotypes is shown in the simulation Section 5.3.1 below.

 \subsection{Model Selection Using Bayes Factor}
Bayes factors are central to the Bayesian model selection and comparison.
The Bayes factor for comparing model $M_0$ and
$M_1$ is defined as: $BF=\frac{P(Y|M_1)}{P(Y|M_0)}$. Assuming equal model priors for $M_0$ and $M_1$, the
posterior odds of the two models equals to the Bayes factor. 
A calibration of the Bayes factor is given by \cite{KassRaftery:1995ys} where 
$\log BF>1$ supports $M_1$ and $\log BF<0$ supports $M_0$. 

The calculation of $P(Y|M_k)$ in our setting is challenging as it involves high dimensional
integration. We follow the procedure of \cite{LeeSong:2002ys} and   use the { parametric arithmetic mean path (PAMP)} implementation of  the path sampling \citep[see also][]{Dutta:2011ys} to compute the Bayes
factor.  Specifically, we assume an unnormalized density function $f_{\theta}$ such that
 $f_{\theta_0}$ and $f_{\theta_1}$ are the sampling densities  for models $M_0$  and  $M_1$, respectively. The two models are connected 
 in the parametric space $\Theta$ via a path 
 ${\cal{P}}=\{ \theta_{g}=g \theta_0+(1-g)\theta_1: g \in[0,1]\}$, where each $\theta_g \in {\cal{P}}$ corresponds to a model $M_g$ for which the sampling density function is $f(\theta_g)$.
 The Bayes factor can then be calculated using the identity 
\begin{equation}\label{eqn:bayes_factor_identity}
\log BF=  \log \frac{P(Y|M_1)}{P(Y|M0)}=\int_0^1\!
E_{\Omega,\Theta}[{\bf \large U}(Y,\Omega,\Theta,g)]dg,
\end{equation}  where
$E_{\Omega,\Theta}$ denotes the expectation with respect to the
density $p(\Omega,\Theta|Y,g)$, and
$\mathbf{U}(Y,\Omega,\Theta,g)=\frac{\partial}{\partial g}\log p(Y,\Omega|\Theta,g).$ The dependence of $p(Y,\Omega|\Theta,g)$ on $g$ is due to $p(Y,\Omega|\Theta,g) \propto p(\Omega|Y,\Theta) f_{\theta_g}(Y|\Theta)$.

To compute the integral in equation (\ref{eqn:bayes_factor_identity}),
we choose $S$ fixed  values $\{g_{1},\ldots, g_{S}\}$ such that 
$g_{(0)}=0<g_{(1)}<g_{(2)}<...<g_{(S)}<1=g_{(S+1)}$ and then estimate
$\log BF$ by
\begin{equation}\label{Bayes_factor_estimate}
\widehat{\log BF}=\frac{1}{2}\sum_{s=0}^S\!(g_{(s+1)}-g_{(s)})(\overline{{\bf \large U}}_{(s+1)}+\overline{{\bf \large U}}_{(s)}),
\end{equation}
where $\overline{{\bf \large U}}_{(s)}$ is the average of the values
${\bf \large U}(Y,\Omega,\Theta,g_{(s)})$ over all the MCMC samples from
$p(\Omega,\Theta|Y,g_{(s)})$. That is, 
\begin{equation}\label{Bayes_factor_U}
\overline{{\bf \large U}}_{(s)}=\frac{1}{M}\sum_{k=1}^M \! \large U(Y, \Omega^{(k)}, \Theta^{(k)}, g_{(s)}),
\end{equation}
in which $\{\Omega^{(k)}, \Theta^{(k)}; k=1,..., M\}$ are the samples draws from $p(\Omega,\Theta|Y,g_{(s)})$.
To estimate $\log BF$, we run the PX$-$HC (or PH$^2-$HC) sampling algorithm for each of the grid points, calculate the values of 
the parameters  and, finally, compute $\overline{\bf \large U}$. This method is also called 
Path Sampling with Parameter Expansion (PS$-$PX) by \cite{Ghosh:2009ys}. 
In the context of pleiotropy studies,  Bayes factors can be applied to different hypotheses. In the remaining part of this section we illustrate two instances of substantial interest.
 
\subsubsection{Selection of Relevant Phenotypes via Bayes Factors.}
We are concerned here with assessing the factor loadings, $\lambda_j$s, as discussed in Section 4.1. 
Suppose that we are interested in testing $\lambda_{j_0}=0$. 
Let $M_g$ be the model with factor loadings 
equal to $(\lambda_1, \ldots,\lambda_{j_0-1},g\lambda_{j_0}, \lambda_{j_0+1}, \ldots,\lambda_J)',$ where $g\in [0,1]$. Thus, model $M_0$ has $(\lambda_1, \ldots, \lambda_{j_0-1}, 0, \lambda_{j_0+1}, \ldots, \lambda_J)'$, and
$M_1$ has  $(\lambda_1, \ldots, \lambda_{j_0-1},\lambda_{j_0}, \lambda_{j_0+1}, \ldots,\lambda_J)'$.
The first part  of the LVM for $M_g$ 
that links the phenotype $Y_j$s and the LV  $U$ is 
\begin{equation}\label{model Mg}
\begin{array}{cc}
V_{citj}=\beta_{0j}+W_{cit}^T\beta_{j}+\lambda_jU_{cit}+b_{cij}+e_{citj},  &\mbox{ if } j\neq j_0\\
V_{citj_0}=\beta_{0j_0}+W_{cit}^T\beta_{j_0}+g\lambda_{j_0}U_{cit}+b_{cij_0}+e_{citj_0}, \\
\end{array}
\end{equation}
where $V_{citj}=Y_{citj}$ for continuous phenotypes and $V_{citj}=\widetilde{Y}^b_{citj}$ for binary phenotypes.
The second part of the LVM remains unchanged. We then have
$${\bf \large U}(Y,\Omega,\Theta,g_{(s)})=\frac{1}{\sigma^2_{j_0}}\sum_{c=1}^C \sum_{i=1}^{N_c}\sum_{t=1}^T \left[V_{citj_0}-(\beta_{0j_0}+W_{cit}^T\beta_{j_0}+g_{(s)}\lambda_{j_0}U_{cit}+b_{cij_0})\right]\lambda_{j_0}U_{cit},$$
where $\sigma^2_{j_0}=1$ for the binary phenotypes. 
The $logBF$ can then be obtained via Equations (\ref{Bayes_factor_estimate}) and (\ref{Bayes_factor_U}) with
$\bf \large U$ calculated using the above equation.

\cite{Dutta:2011ys} showed that the calculation of Bayes factor is valid when
$\lim_{g\rightarrow 0} \!E_{\Omega,\Theta}[{\bf \large U}(Y,\Omega,\Theta,g)]$ is finite. The condition holds when the prior distribution for $\lambda$ has the first two moments finite, which is true for our   relatively diffuse folded-t prior with ten df. To explore the sensitivity of the Bayes factor estimation to the changes of df in the folded-t prior, we also consider df $\in \{3, 10, 40, 90\}$.  

Another concern is the tuning of the grid sizes when using equation (\ref{Bayes_factor_estimate}) as the approximation of $\log BF$. \cite{Dutta:2011ys} suggested a grid size no bigger than 0.01, corresponding to 100 grid numbers in $[0,1]$. However, due to conflict between the prior and sampling distributions, 
when $\lambda$ is large,  $E_{\Omega,\Theta}[{\bf \large U}(Y,\Omega,\Theta,g)]$ has large spikes when $g$ is close to zero, but it stabilizes gradually as  $g$ increases. Thus we suggest to use uneven grid size scheme so that smaller grid sizes are chosen for $g$ near zero. For example, in our simulations below
 we set the grid number to be 15 for $g\in [0,0.1]$ and also 15 for $g \in [0.9,1]$ and observed good results.
  
\subsubsection{Selection of Pleiotropic Genetic Marker via Bayes Factors}
An important inferential focus in genetic pleiotropy study is the selection of genetic marker(s) with pleiotropic effect.
Particularly, we are interested in testing the effect  of a genetic marker on the LV.  Suppose that the fixed-effect covariates $X$ in the second part of the LVM has two components where $X_{1}$ is the set of clinical covariates and $X_{2}$ is the genotype of the marker of interest.
The two competing LV models are then
 \[
 M_0:\left\{
\begin{array}{ll}
 y^c_{citj}=\beta_{0j}+W_{cit}^T\beta_{j}+\lambda_jU_{cit}+b_{cij}+e_{citj}, &j=1,...,J_1,\\
 \widetilde{y}^b_{citj}=\beta_{0j}+W_{cit}^T\beta_{j}+\lambda_jU_{cit}+b_{cij}+e_{citj},  & j=J_1+1,...,J,\\
U_{cit}=X_{1_{cit}}^T\alpha_1+Z_{cit}^T a_c+Q_{cit}^T d_{ci}.&\\
\end{array}
\right .
\]
\[
 M_1:\left\{
\begin{array}{ll}
 y^c_{citj}=\beta_{0j}+W_{cit}^T\beta_{j}+\lambda_jU_{cit}+b_{cij}+e_{citj}, &j=1,...,J_1,\\
  \widetilde{y}^b_{citj}=\beta_{0j}+W_{cit}^T\beta_{j}+\lambda_jU_{cit}+b_{cij}+e_{citj},  & j=J_1+1,...,J,\\
U_{cit}=X_{1_{cit}}^T\alpha_1+X_{2_{cit}}^T\alpha_2+Z_{cit}^T a_c+Q_{cit}^T d_{ci}.&\\
\end{array}
\right .
\]
The two models can be linked up by the parameter $g \in [0,1]$ as:
\[
 M_{g}:\left\{
\begin{array}{ll}
 y^c_{citj}=\beta_{0j}+W_{cit}^T\beta_{j}+\lambda_jU_{cit}+b_{cij}+e_{citj}, &j=1,...,J_1,\\
  \widetilde{y}^b_{citj}=\beta_{0j}+W_{cit}^T\beta_{j}+\lambda_jU_{cit}+b_{cij}+e_{citj},  & j=J_1+1,...,J,\\
U_{cit}=X_{1_{cit}}^T\alpha_1+g X_{2_{cit}}^T\alpha_2+Z_{cit}^T a_c+Q_{cit}^T d_{ci}.&\\
\end{array}
\right .
\]
Let $L_{g}$ be the complete-data likelihood for model $M_{g}$,
then
$${\bf \large U}(Y,\Omega,\Theta,g_{(s)})=\frac{\partial  \log L_{g}}{\partial g}
= \sum_{c=1}^C \sum_{i=1}^{N_c}\sum_{t=1}^T
 \left(U_{cit}-X_{1_{cit}}^T\alpha_1-g_{(s)} X_{2_{cit}}^T\alpha_2-Z_{cit}^T a_c-Q_{cit}^T d_{ci}\right)
      X_{2_{cit}}^T\alpha_2.$$
%
Again $\log BF$ can be obtained using (\ref{Bayes_factor_estimate}) and (\ref{Bayes_factor_U}) with
$\bf \large U$ calculated as above. 

\section{Simulation study} \label{section pleiotropy simulation}

We have identified three practically relevant issues related to the model's performance that we wanted to study via simulations: 1) the performance of our proposed method in parameter estimation, 2) the effect of ignoring the family structure, and 3) the performance of the proposed  variable selection methods introduced in Section 4. 

Table  \ref{tb_simulation} provides the details of the simulation models in terms of phenotype $Y$ and covariates $W$ and $X$ specifications.
Without loss of the generality, in all simulations we assume that there are 500 families/clusters, each contributing 1, 2, 3, 4 or 5  siblings with probability, respectively, 0.3, 0.4, 0.2, 0.07 and 0.03. To simulate the genotypes for the siblings, we set the minor allele frequency (MAF) to be 0.1, and we assume that the parental genotypes for the 500 families  follow Hardy Weinberg equilibrium(HWE). The siblings' genotypes are then obtained by using Mendel's first law of segregation. The parental genotypes however are not used for the analysis because they are often not available in practical settings. Continuous covariates are assumed to follow a $N(0,1)$ distribution with correlation coefficient within the family (familial correlation) being 0.3 and  individual-specific trajectories for the covariate (serial correlation) follow an AR(1) model with autocorrelation being 0.3. We also assume that each phenotype is measured over time five times.
Unless specified otherwise, the default choice is $\pi_j\sim \text{Beta}(1,1)$ in the spike-and-slab prior given in \eqref{lambda-prior}, assuming no prior information on the relationship between phenotype $j$ and the LV.  In all the simulations, we run the MCMC algorithm for 25,000 iterations, discarding the first 5,000 samples as burn-in.

\subsection{Parameter estimation for general traits.} \label{section pleiotropy simulation 1}
      
To evaluate the performance of the proposed parameter estimation, we assume here that five phenotypes are of interest among which $ y_1$, $y_2$ and $y_3$ are continuous,
 and $y_4$ and $y_5$ are binary (Tables \ref{tb_simulation} and  \ref{tb_pleiotropy_sim1}). 
 We use a total of 100 replications to study the
variation of estimates and also compare the
performance of our PX$^{2}-$HC algorithm
 with those of the standard algorithms.  
 
Figure 1 shows that, compared to AC-PX-HC, the standard PX algorithm with the DA scheme of \cite{AlbertChib:1993ys},  the proposed PX$^{2}-$HC algorithm drastically reduces the autocorrelations between the chains's realizations which implies an increase in Monte Carlo efficiency. The improvement is also evident in Figure 2, when comparing the 
standard Gibbs 
(left column) with the proposed PX$^{2}-$HC scheme (right plots). In the Appendix,  we provide the corresponding trace plots to illustrate the improved mixing of the PX$^{2}-$HC chain.
 
Table \ref{tb_pleiotropy_sim1} shows the true values, mean estimates  and root mean square errors (RMSE) for all the parameters in the model. Results show that  the estimates have good accuracy with small
bias and RMSE. However,  parameter estimation for binary responses is less accurate than for continuous phenotypes, which is not surprising given the discrepancy between the information provided by the two types of data. 

Figure \ref{HPDI fig} shows that around $95\%$ of the HpdIs
 for the factor loadings $\lambda$s cover the true values.  However,
 this remains to be validated theoretically for more general setting as results seem to suggest a matching prior type of result.
 
\subsection{Effect of Ignoring Family Structure}\label{section pleiotropy simulation 2}
In this simulation, we only consider the  three
continuous phenotypes $y_1$, $y_2$ and $y_3$ and generate data with familial and serial correlations as  above in Section 5.1.
We compare parameter estimates obtained using the proposed method to model the familial correlation
with the estimates obtained using the method of \cite{RoyLin:2000ys} assuming samples are independent of each other.

The simulation results in Table
\ref{tb_pleiotropy_sim2} show that failure to account for the familial correlation present in the data
will not only yield incorrect conclusion of the serial correlations between phenotypes $\Sigma_D$,  but also cause over-estimation of the factor loading $\lambda_j$, which quantifies the strength of the relationship between phenotype $y_j$ and the latent variable $U$,  and under-estimation of  $\alpha_1$  and $\alpha_2$, which respectively,  represents the effect of the clinical covariate and genetic marker on  $U$.

   \subsection{Model and Variable Selection} \label{section pleiotropy simulation 3}
\subsubsection{Selection of Relevant Phenotypes via Spike-and-Slab Priors}
\label{section pleiotropy simulation 3.1}
 To assess the performance of phenotype selection using the spike-and-slab prior \eqref{lambda-prior} as described in Section 4.1,
 we consider here four continuous phenotypes $y^c$ with factor loading $\lambda_j$s being set to
 $(0.5, 0.05, 0.02, 0)$ and three binary phenotypes $y^b$ with $\lambda_j$s being $(0.2, 0.01, 0)$.
 We chose these values so that   the strength of the association between a phenotype and the LV $U$ ranges from strong ($\lambda_j=0.5$) to no association ($\lambda_j=0$). We set $\pi\sim \text{Beta}(0.25,1)$ which favours {\it a priori} the null ($\lambda_j=0$).  After the calculation of the posterior frequency of $\omega_j=1$ for each phenotype in the 100  replicates, a threshold of $\phi=50\%$ is used on the posterior probability $Pr(\lambda_j> 0|\boldsymbol Y)$.

Simulation results show that the type I error of this selection procedure is about 5\% in that, for the phenotype not associated with the latent disease status ($\lambda_j=0$), around 5\% of the replications have $Pr(\lambda_j> 0|\boldsymbol Y) \geq 0.5$. For phenotypes moderately or strongly associated the latent variable ($\lambda_j=0.05$ or higher), we have 100\% power of making the correct decision. However, power is reduced for weakly associated phenotypes as expected. Specifically for the case consider here, power is 71\% for the phenotype with $\lambda_j=0.02$ and 16\% for $\lambda_j=0.01$.

\subsubsection{Selection of Relevant Phenotypes via Bayes Factors} \label{section pleiotropy simulation 3.2}

We have also examined the performance of phenotype selection via Bayes factors {for the comparison of model $M_1$ (assuming $\lambda>0$) and model $M_0$ (assuming $\lambda=0$)} using the same simulation model as in the previous section. The $logBF$s for the $\lambda$'s are calculated using the folded-t prior with df $=10$  as described in Section 4.2.1.
   
Results in Table \ref{tb_sim3.2} show that when phenotype $y_j$ is truly associated with the latent variable  with $\lambda_j \geq 0.05$, the average estimated $log(BF) > 5.9$. This and combined with the SD shown in the table suggest that, for such cases, the Bayes factor criterion chooses the correct alternative model 
$100\%$ of the times. When phenotype $y_j$ is truly not associated with the LV ($\lambda_j=0$), the Bayes factor criterion correctly favors the null model with the average estimated $log(BF) < -2.4$.
However, the Bayes factor criterion has little ability in identifying weakly associated phenotypes ($\lambda_j\leq 0.02$), which is inferior to the spike-and-slab approach in the previous section. We have also explored the use of other folded-t priors with  df ranging from 3, 10, 40 to 90. We find that all the prior settings give us the same conclusion on the model selection.  

\subsubsection{Selection of Pleiotropic Genetic Marker or Other Covariates}  \label{section pleiotropy simulation 3.3}
For this purpose, we assume that there are five indirect fixed-effect covariates $X_{1},\ldots,X_{5}$ in the second part of the LVM that models the effect of the genetic marker and other covariates on the latent variable. We assume that $X_3$  is the genotype of the marker of interest and the remaining ones are  standardized continuous variables. Coefficients
$(\alpha_1,\alpha_2,\alpha_3,\alpha_4,\alpha_5)$ quantify the effects of the five covariates which are set to be $(1.0, -0.5, 0.2, 0, 0)$. We also assume that there are two continuous standardized direct fixed-effect covariates $W_1$ and $W_2$ in the first part of the LVM with effects ($\beta_1=0.5$ and $\beta_2=0.3$)  on all the phenotypes.

Table \ref{tb_sim3.3} shows the estimated $log(BF)$ for comparing the null model assuming $\alpha_j=0$ with the alternative of association. The prior distribution for $\alpha$ in \eqref{eqn:latent} is $\alpha\sim N(0,\mathbf{I}_5)$. Results show that the proposed Bayes factor criterion has the ability to detect the association between the genetic marker and the latent variable, therefore pleiotropy, with the average estimated $log(BF) = 6.3$ and SD $=2.39$. The Bayes factor variable selection criterion also has good result  for the two associated covariates $X_1$ and $X_2$ for which the average $log(BF)$ estimate is, respectively, 337.5 and 6.3, as well as  for the two covariates $X_4$ and $X_5$ with zero effect, for which the estimated $log(BF)$ is consistently less than $-1.9$ with  SD $=0.42$. 

\section{Application to a genetic study of type 1 diabetes (T1D) complications.}\label{application}

Here we demonstrate the practical utility of the proposed LVM method by investigating the blood pressure data from a genome-wide association study (GAWS) of various T1D complications \citep{Paterson:2010ys}. 
The study sample consists of $n=1300$ individuals with T1D from the Diabetes Control and Complications Trial (DCCT). Various phenotypes thought be to related to T1D complication severity, including glycosylated hemoglobin (HbA1c) and diastolic  (DBP) and systolic blood pressure (SBP), were collected from each subject over the course of the DCCT. Additional covariates such as sex and body mass index (BMI) were also collected, and individuals were from  two different cohorts and subjected to two treatment types (conventional vs.\ intensive). Over 800K SNPs  were genotyped by the Illumina 1M bead chip assay for these individuals. 

Because T1D is a complex disease with various complication measures (the observed phenotypes), it is of great interest to quantify the conceptual latent complication status, as well as to understand the influencing factors (both genetic markers and clinical covariates). In addition, it is valuable to determine if the various observed phenotypes are  truly associated with the latent variable. However, due to lack of suitable statistical methodology, previous analyses have been limited to the standard uni-phenotype approach, analyzing one phenotype at a time.  
For example, \cite{ye} recently performed GAWS,  {\it separately},  for  DBP and SBP,
and they identified  rs7842868 on chromosome 8 as a SNP significantly associated with DBP.

Our goal here is to formally perform a multi-phenotype analysis, jointly analyzing DBP and SBP using the proposed Bayesian LVM methodology. This approach allows us not only to determine if rs7842868 is associated with  the latent conceptual T1D complication variable, but also to test if DBP and SBP are truly related to the LV. 
 It is also of practical interest to study whether there are other phenotypes such as hyperglycemia (HPG) influence the latent complication severity variable.

In our application, we investigate three phenotypes $Y$ among which two are continuous (DBP and SBP)  and one is binary (HPG $=1$ for hyperglycemia if and $=0$ for normal glycemia), all are longitudinal. Hyperglycemia at a given time point is defined if the corresponding HbA1c is greater or equal to 8. 
Among the available covariates, based  on suggestions from clinicians, covariates $W$ that have direct effects on the phenotypes include BMI, and covariates $X$ that have direct effects on the LV include sex, cohort and treatment.  Among the 10 longitudinal measurements available, there are  significant amount of missing data  after the $7^{th}$ measure  (due to staggered
 entry) while there are little missing data before the $5^{th}$ measure.
Therefore, we only use the first five measurements. We treat the remaining missing data as Missing at Random (MAR), and we replace the missing data with the means of all the other  measurements. 
In this dataset, there is only one person in each family therefore there is no familial correlation, but the proposed methodology remains suitable by assuming the cluster size being 1. 

We first consider  rs7842868, a SNP found by \cite{ye}  to be associated with DBP. In this case $X_1$ is the genotype of rs7842868. Results in Table \ref{tb_application} show that DBP and SBP are clearly associated with the latent variable  with estimated $logBF$ over 100, while HPG is not. We also applied the spike-and-slab prior method for phenotype selection as discussed in Sections 4.1 and 5.3.1, the poster probability 
$Pr(\lambda_j> 0|\boldsymbol Y)$ is 1 for DBP and SBP and 0.235 for HPG. Thus all model selection criteria consistently suggest that both DBP and SBP are significantly  related to the latent variable but not HPG. 
Results also show that SNP rs7842868 is significantly associated with the latent variable with estimated $logBF$ over 10 and the 95\% HpdI not covering 0. The sign of the effect suggest that the minor allele of the SNP is protective in that it decreases the latent complication severity score. The combined evidence from both parts of the LVM show that rs7842868 has pleiotropic effect on the two blood pressures.
Sex and cohort are also found to be significantly associated with the latent variable but not treatment.

We then investigate  rs1358030, a SNP found by \cite{Paterson:2010ys} to be associated with HbA1c. In this case, $X_1$ is the genotype of rs1358030. Based on results in Table \ref{tb_application}, there  is no evidence that rs1358030 is significantly associated with the latent  variable. Our application here considers the binary hyperglycaemia (HPG) as the third phenotype of interest instead of the continuous HbA1c variable. Besides clinical consideration, this choice also allows us to evaluate the proposed method for general traits as described in Section 3.2.2


To further evaluate the proposed method, we simulate genotypes for two NULL SNPs that are not associated with the phenotypes of interest. One SNP has MAF equal to 0.25, the MAF of rs7842868), and the other one has MAF equal to 0.35, the MAF of  rs1358030. As expected, no significant associations are detected.

 \vspace{-10pt}
\section{Conclusion and Disucssion}

{
We propose here a Bayesian latent variable approach to joint model multiple outcomes, motivated by genetic association studies of pleiotropic effects. The method can handle continuous and binary responses while accounting for  serial and familial correlation structures in the data. The postulated latent variable represents the underlying severity or complication level of a trait and characterizes the totality of multiple observed phenotypes of interest. If additional phenotypes were to be measured, the latent variable could change its significance since it would encapsulate a richer set of manifest variables. The central feature of the model is that it allows us to consider the strength of  dependence  between the genotype and each of the phenotypes in a unified manner. The Bayesian approach  takes into account all the uncertainty present in the model and incorporates prior information if available. The computational challenges are met via the use of parametric expansion techniques. 

An important issue for pleiotropy studies is the assessment of importance of each variable. We have adopted two Bayesian techniques that are shown to be effective in variable selection. We found that both Bayes factor and
spike-and-slab prior perform well with the latter slightly more efficient in terms of detecting weak signals.


Our proof-of-principle application to a genetic study of type 1 diabetes complications demonstrates the utility of the method in a real data setting. So far, genetic association studies of various T1D complication-related measures have been limited to studying one phenotype at a time. The proposed method jointly analyzes two continuous and one binary phenotypes of interest,  and it provides evidence for the association between the phenotypes and the latent severity of T1D complication, the association between the latent variable and genetic markers of interest, and the effect of other covariates on the phenotypes and the latent variable. 

The computational load in the current implementation of the proposed method makes it impractical to perform a genome-wide search for pleiotropic genetic variants. The recent advances in parallel computing can partially alleviate this constraint. Alternatively, a two-stage approach can be used in which a simple and less stringent selection procedure is first used to select a moderate number of candidate variants for further investigation using the proposed method. The uncertainty inherited from the first-stage selection, however, must be accounted for in the models used in the second stage. 
}

\section*{Acknowledgments}
 This work was supported by the Natural Sciences and Engineering Research Council (NSERC) of Canada  to RVC, NSERC (250053-2008) to LS,  and the Canadian Institutes of Health Research (CIHR; MOP 84287) to RVC and LS, the Ontario Graduate Scholarship (OGS) to LX. A.D.P. holds a Canada Research Chair in the
Genetics of Complex Diseases and received funding from Genome Canada through the Ontario Genomics Institute. 
 
The DCCT/EDIC Research Group is sponsored through research contracts from the National Institute of Diabetes, Endocrinology and Metabolic Diseases of the National Institute of Diabetes and Digestive and Kidney Diseases (NIDDK) and the National Institutes of Health (N01-DK-6-2204, R01-DK-077510). The Diabetes Control and Complications Trial (DCCT) and its follow-up the Epidemiology of Diabetes Interventions and Complications (EDIC) study were conducted by the DCCT/EDIC Research Group and supported by National Institute of Health grants and contracts and by the General Clinical
Research Center Program, NCRR. The data and samples from the DCCT/EDIC
study were supplied by the NIDDK Central Repositories. This manuscript was
not prepared under the auspices of and does not represent analyses or
conclusions of the NIDDK Central Repositories, or the NIH.
 
\bibliographystyle{ims}


\begin{landscape}
\begin{table}[h!] 
\caption{Simulation and Application Models } \label{tb_simulation} {
\addtolength{\tabcolsep}{-5pt}

\begin{center}
               \begin{tabular}{l|l|l|l|l|l} \hline \hline            
\hspace{1cm} &	\multicolumn{4}{c|}{\bf Simulation Model Considered in}	& Application Model\\			
&	\: Section 5.1&	\: Section 5.2	&\: Sections 5.3.1\&5.3.2\: \: 	 &	\: Section 5.3.3 & \: \: \: Section 6\\ \hline\hline
\multicolumn{5}{l}{Phenotypes of interest}\\	\hline				
$y_1$&	$\surd$ (continuous)&$\surd$ (continuous)&$\surd$ (continuous)&	$\surd$  (continuous) & SBP (continuous)\\
$y_2$&	$\surd$ (continuous)&$\surd$ (continuous)&$\surd$ (continuous)&	$\surd$  (continuous) & DBP (continuous)\\
$y_3$&	$\surd$ (continuous)&$\surd$ (continuous)&$\surd$ (continuous)&	$\surd$  (continuous) & HPG (binary)\\
$y_4$&	$\surd$ (binary)        &	                              &$\surd$ (continuous)&	$\surd$ (binary) \\
$y_5$&	$\surd$ (binary)        &	                              &$\surd$ (binary)         & $\surd$ (binary) \\
$y_6$&	                                     &	                              &$\surd$ (binary)         & \\
$y_7$&	                                     &	                              &$\surd$ (binary)         & \\ \hline
\multicolumn{6}{l}{Covariates with direct effect $\boldsymbol \beta$ on $Y$}\\		\hline			
$W_1$&	$\surd$ { (continuous)}         &$\surd$  { (continuous)}       &$\surd$  { (continuous)}                  &	$\surd${ (continuous)}    & BMI\\
$W_2$&	                                    &	                             &	                                 &	$\surd${ (continuous)}    & \\ \hline
\multicolumn{6}{l}{Covariate with indirect effect on $Y$ but with direct effect $\boldsymbol \alpha$ on the latent variable $U$}	\\		\hline	
$X_1$&	$\surd$ {(continuous)}     &	$\surd$ {(continuous)}    &$\surd${ (continuous)}                            &	$\surd${ (continuous)}    & genotype\\
$X_2$&	$\surd$ (genotype)&	$\surd$ (genotype)   &$\surd$ (genotype)     &	$\surd${ (continuous)}    & sex\\
$X_3$&	                                  &	                                     &	                                &	$\surd$ (genotype)& cohort\\
$X_4$&	                                  &	                                     &	                               &	$\surd$ {(continuous)}    & treatment\\
$X_5$&	                                  &	                                     &	                               &	$\surd$ {(continuous)}    & \\ \hline
\multicolumn{6}{l}{Latent variable $U$ has direct effect $ \boldsymbol \lambda$ (also called factor loading) on each of the $Y$s}	\\\hline\hline	 
\end{tabular}
\end{center}}
\end{table}	
	
\end{landscape}
	
\begin{table}[h!] 
\caption{\em Performance of the proposed parameter estimation method. First three phenotypes $y_1$, $y_2$ and $y_3$ are continuous and the last two phenotypes $y_4$ and $y_5$ are binary.
Results are based on 100 replicates that were simulated as described in Section 5.1 and Table \ref{tb_simulation}.}\label{tb_pleiotropy_sim1}{
\addtolength{\tabcolsep}{-5pt}

\begin{center}
               \begin{tabular}{c|c|c|l} \hline \hline
\multicolumn{4}{c}{\bf First part of the latent variable model} \\ 	\hline \hline
Parameter \: &	 True value \: \:  & \: \: Estimate \: \:   &\hspace{1cm}RMSE\\ \hline 
	 \multicolumn{4}{l}{\hspace{0.8cm}$\lambda_j$, the factor loading for phenotype $y_j$ and the latent variable $U$} \\ \hline
$\lambda_1$ &	1.0	&1.000	&\hspace{1cm}0.007\\
$\lambda_2$&	1.0	&1.000	&\hspace{1cm}0.007\\
$\lambda_3$&	1.0	&1.000	&\hspace{1cm}0.007\\
$\lambda_4$&	1.0	&1.000	&\hspace{1cm}0.022\\
$\lambda_5$&	1.0	&0.998	&\hspace{1cm}0.025\\ \hline
	  \multicolumn{4}{l}{\hspace{0.8cm}$\beta_j$ for phenotype $y_j$ and covariate $W$} \\ \hline
$\beta_1$ &	1.0	&1.000	&\hspace{1cm}0.011\\
$\beta_2$&	1.0	&1.000	&\hspace{1cm}0.011\\
$\beta_3$&	1.0	&1.000	&\hspace{1cm}0.011\\
$\beta_4$&	1.0	&1.005	&\hspace{1cm}0.026\\
$\beta_5$&	1.0	&0.996	&\hspace{1cm}0.031\\ \hline \hline
	 \multicolumn{4}{c}{\bf Second part of the latent variable model} \\ 	\hline \hline
Parameter \: 	 &	True value &	Estimate &\hspace{1cm}RMSE\\ \hline 
	  \multicolumn{4}{l}{\hspace{0.8cm}$\alpha$ for the latent variable $U$ and covariates ($\alpha_2$ for the genetic marker $X_2$)} \\ \hline
$\alpha_1$ &	1.0	&0.999	&\hspace{1cm}0.011\\
$\alpha_2$ &	1.0	&1.011	&\hspace{1cm}0.074\\ \hline\hline
\multicolumn{4}{c}{\bf Correlation parameters} \\ 	\hline \hline
Parameter \: &	True value &	Estimate &\hspace{1cm}RMSE\\ \hline 
$\tau_1^2$ &	0.2	&0.200	&\hspace{1cm}0.014\\
$\tau_2^2$&	0.2	&0.201	&\hspace{1cm}0.012\\
$\tau_3^2$&	0.2	&0.200	&\hspace{1cm}0.013\\
$\tau_4^2$&	0.2	&0.202	&\hspace{1cm}0.033\\
$\tau_5^2$&	0.2	&0.200	&\hspace{1cm}0.032\\ 
$\Sigma_A$&	0.5	&0.519	&\hspace{1cm}0.046\\
$\Sigma_D$&	0.3	&0.320	&\hspace{1cm}0.032\\ \hline \hline
	 
\end{tabular}
\end{center}}
\end{table}

\begin{table} \caption{Effect of ignoring familial correlation present in the data. Results are  the averages over 100 simulation replications. The coefficient $\alpha_2$ evaluates the effect of a genetic marker on the latent variable $U$ 
($\alpha_1$ for a clinical covariate on $U$), and $\lambda_j$s evaluate the effect of $U$ on the three phenotypes of interest.
Details of the simulations are in Section 5.2 and Table \ref{tb_simulation}.} \label{tb_pleiotropy_sim2}{
\addtolength{\tabcolsep}{-5pt}

 \begin{center}
               \begin{tabular}{c|c|c||c|c|c||c|c|c}
               \hline \hline
         \multicolumn{2}{c|}{ }& &\multicolumn{3}{|c||}{With Familial Corr.}&
          \multicolumn{3}{c}{Without Familial Corr.}\\
         \cline{4-9}
        \multicolumn{2}{c|}{Parameters}&True Value&\: \: Bias \:\:  &\: SD \: \:  &\: \: RMSE \: \: &\: \: Bias \: \:  &\: \: SD \: \: &\: \: RMSE \: \: \\
                        \hline \hline
   \multirow{3}{*}{$\beta$}&$\beta_{1}$&1.0&-0.002&0.018&0.019&-0.003&0.028&0.027\\
             & $\beta_{2}$&1.0&-0.001&0.020&$0.019$&-0.002&0.024&0.028\\
           & $\beta_{3}$&1.0&-0.002&0.019&$0.019$&-0.003&0.027&0.028\\
               \hline
      \multirow{2}{*}{$\alpha$}&  $\alpha_1$&1.0&0.009&0.074&0.075&\bf -0.233&0.099&\bf 0.235\\
                           &$\alpha_2$&1.0&0.009&0.020&0.021& \bf -0.227&0.028 &\bf 0.244\\

\hline
  \multirow{3}{*}{$\lambda$}&$\lambda_1$&1.0&-0.009&0.014&0.015&\bf 0.308& 0.029&\bf 0.310\\
             & $\lambda_2$&1.0&-0.009&0.014&0.016&\bf 0.308&0.028&\bf 0.309\\
          & $\lambda_3$&1.0& -0.009&0.015&0.015&\bf 0.308&0.029&\bf 0.309\\
 \hline
  \multirow{3}{*}{$\tau^2$}&$\tau^2_1$&0.2&-0.002&0.013&0.015&-0.001&0.015&0.015\\
             & $\tau^2_2$&0.2& -0.001&0.012&0.014& -0.001&0.013&0.015\\
      &$\tau^2_3$&0.2&-0.002&0.012&0.013& 0.000&0.014&0.014\\
\hline
  \multirow{3}{*}{$\sigma^2$}&$\sigma^2_1$&0.1& 0.002&0.004&0.004&0.002&0.004&0.004\\
             & $\sigma^2_2$&0.1&0.001&0.004&0.004&0.001&0.004&0.004\\
           & $\sigma^2_3$&0.1& 0.002&0.004&0.004&0.002&0.004&0.004\\

\hline
      \multirow{3}{*}{$\Sigma_A$}&$(\Sigma_{A})_{11}$&1.0&-0.003&0.08&0.089&N/A&N/A&N/A\\
             & $(\Sigma_{A})_{12}$&0.0& 0.004&0.06&0.064&N/A&N/A&N/A\\
           & $(\Sigma_{A})_{22}$&1.0&0.034&0.08&0.085&N/A&N/A&N/A\\
    \hline
 \multirow{3}{*}{$\Sigma_D$}&$(\Sigma_{D})_{11}$&0.1&0.049&0.020&0.053&\bf 0.69&0.06&\bf 0.696\\
             & $(\Sigma_{D})_{12}$&0.0&0.000&0.011&0.011&-0.002&0.022&0.023\\
           & $(\Sigma_{D})_{22}$&0.1&0.029&0.023&0.033&\bf 0.047&0.021&\bf 0.052\\
    \hline \hline

                         \end{tabular}

              \end{center}}
          \end{table}

\begin{table} \caption{The estimated $logBF$ for testing the factor loading $\lambda_j$. $\lambda_j$ quantifies the association between phenotype $y_j$ and the latent variable.  Results are based on 50 simulated replicates.
Details of the simulations are in Section 5.3.2 and Table \ref{tb_simulation}.} \label{tb_sim3.2}{
\addtolength{\tabcolsep}{-5pt}
 \begin{center}
               \begin{tabular}{c|c|c|c|c|c|c|c}
               \hline \hline
               &\multicolumn{4}{c}{ \: Continuous Phenotypes \: }&\multicolumn{3}{|c}{ \: Binary Phenotypes \: }\\
 \hline
            & $y_1$ &  $y_2$ &  $y_3$ &  $y_4$ & $y_5$ & $y_6$ & $y_7$   \\ \hline \hline
              True $\lambda_j$ &0.5&0.05&0.02&0&0.2&0.01&0\\
 \hline
               $\widehat{logBF}$& 121.60&5.91& -1.76&-2.67&22.11&-2.43&-2.43\\
\hline    SD    &6.47& 2.22&1.05&0.47& 8.19& 0.32& 0.13 \\
\hline \hline
\end{tabular}
\end{center}}
\end{table} 

\begin{table} \caption{The estimated $logBF$ for testing  $\alpha=0$. $\alpha$ quantifies the association between the  covariates and the latent variable.  Results are based on 50 simulated replicates.
Details of the simulations are in Section 5.3.3 and Table \ref{tb_simulation}. } \label{tb_sim3.3}{
\addtolength{\tabcolsep}{-5pt}
 \begin{center}
               \begin{tabular}{c|c|c|c|c|c}
               \hline
         &\multicolumn{5}{c}{Covariates with effect on the latent variable}\\
          & $X_1$ &  $X_2$ &  $X_3$ (genotype) &  $X_4$ & $X_5$   \\ \hline \hline
              True $\alpha$ &1.0&-0.5&0.2&0&0\\
 \hline
            $\widehat{logBF}$ & 337.53& 94.99&6.30&-1.94& -1.95\\ 
\hline
             SD&33.25& 11.19& 2.39& 0.46& 0.42\\ 
\hline \hline

\end{tabular}
\end{center}}
\end{table}  

 \begin{table} \caption{Application results.  SNP rs7842868 was previously identified to be associated with diastolic blood pressure (DBP) and SNP rs1358030 was previously identified to be associated with HbA1c. Phenotypes of interest are DBP and systolic blood pressure (SBP), two continuous outcomes, and hyperglycemia (HPG, defined as HbA1c greater or equal to 8), a binary outcome. All phenotypes are thought to be related to type 1 diabetes complication severity.  The coefficient $\lambda$s assess the association between the phenotypes and the latent T1D complication status,  and $\alpha$s evaluate the association between the latent variable and the genetic marker and the other covariates of interest. See Section 6 and Table 1 for more details.} \label {tb_application}{
\addtolength{\tabcolsep}{-5pt}

\begin{center}
    \begin{tabular}{ccccc}
\hline \hline
\multicolumn{5}{c}{\bf Analysis of SNP rs7842868} \\ \hline
& Parameter\: \: & \: \: Estimate \: \: & 95\% HpdI &  $\widehat{logBF}$	\\\hline \hline	

SBP &$\lambda_1$	 &6.621 & (6.153, 7.077) &	114.85 	  \\
DBP& 	$\lambda_2$	 &3.842 &(3.566, 4.110) &	112.98 \\ 
HPG	 &$\lambda_3$& 0.011 &($2.189\times10^{-7}$, $2.975 \times 10^{-2}$) &	-1.05 	\\ \hline
		
rs7842868 &$\alpha_1$&	-0.269 &(-0.372, -0.164)& 	10.06 \\
sex 	&$\alpha_2$ &-0.721& (-0.866, -0.584)& 	62.27  \\
cohort &	$\alpha_3$ &0.443 &( 0.299, 0.585) &	20.15 \\
treatment & $\alpha_4$&	0.128 &(-0.004, 0.263)& 	0.366  \\ \hline \hline
\multicolumn{5}{c}{\bf Analysis of SNP rs1358030} \\ \hline 
 & Parameter& Estimate& 95\% HpdI &  $\widehat{logBF}$	\\\hline \hline	

SBP &$\lambda_1$	& 6.868 &(6.439, 7.302) & 	128.3  \\
DBP& $\lambda_2$ &3.706 & (3.491, 3.933) & 120.2	 \\ 
HPG	 &$\lambda_3$&0.010 &($2.566\times10^{-7}$, $2.740\times10^{-7}$) &-1.034		\\ \hline

rs1358030 & $\alpha_1$&	- 0.039& (-0.049, 0.122)& 	-1.104 \\ 
sex 	&$\alpha_2$ &-0.758 &(-0.880, -0.623) &	64.86  \\
cohort &$\alpha_3$	&0.393 &(0.258, 0.532) &	18.17  \\
treatment 	&$\alpha_4$ &0.088 &(-0.041, 0.220) &	-0.18  \\ \hline \hline
\end{tabular}
\end{center}}

\end{table}  
  
 \begin{figure}
\caption{{\em {Autocorrelation functions (ACF) for the AC-PX-HC (dashed line) scheme 
 and the proposed $PX^2$-HC sampling scheme (solid line),  averaged over 100 simulation replicates.    
 The autocorrelation is based on the posteriors of the factor loadings $\lambda_4$ (left panel) and $\lambda_5$ (right panel) 
for the two binary phenotypes as simulated in Section 5.1 and described in Table 1. }
}}

\includegraphics[trim=0cm 5cm 0cm 3cm, clip=true,height=3.5in,width=0.45\textwidth]{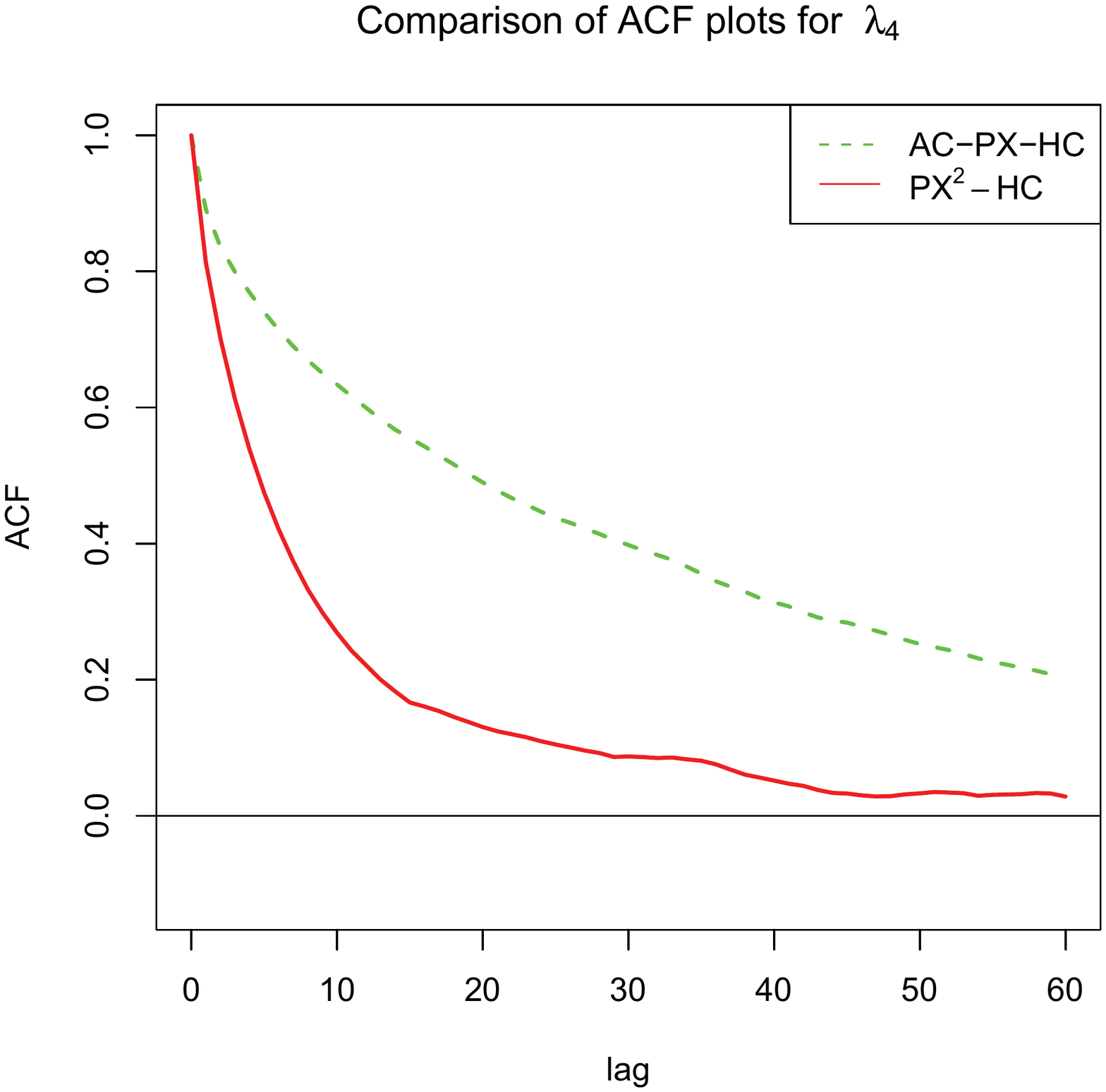}
\includegraphics[trim=0cm 5cm 0cm 3cm, clip=true,height=3.5in,width=0.45\textwidth]{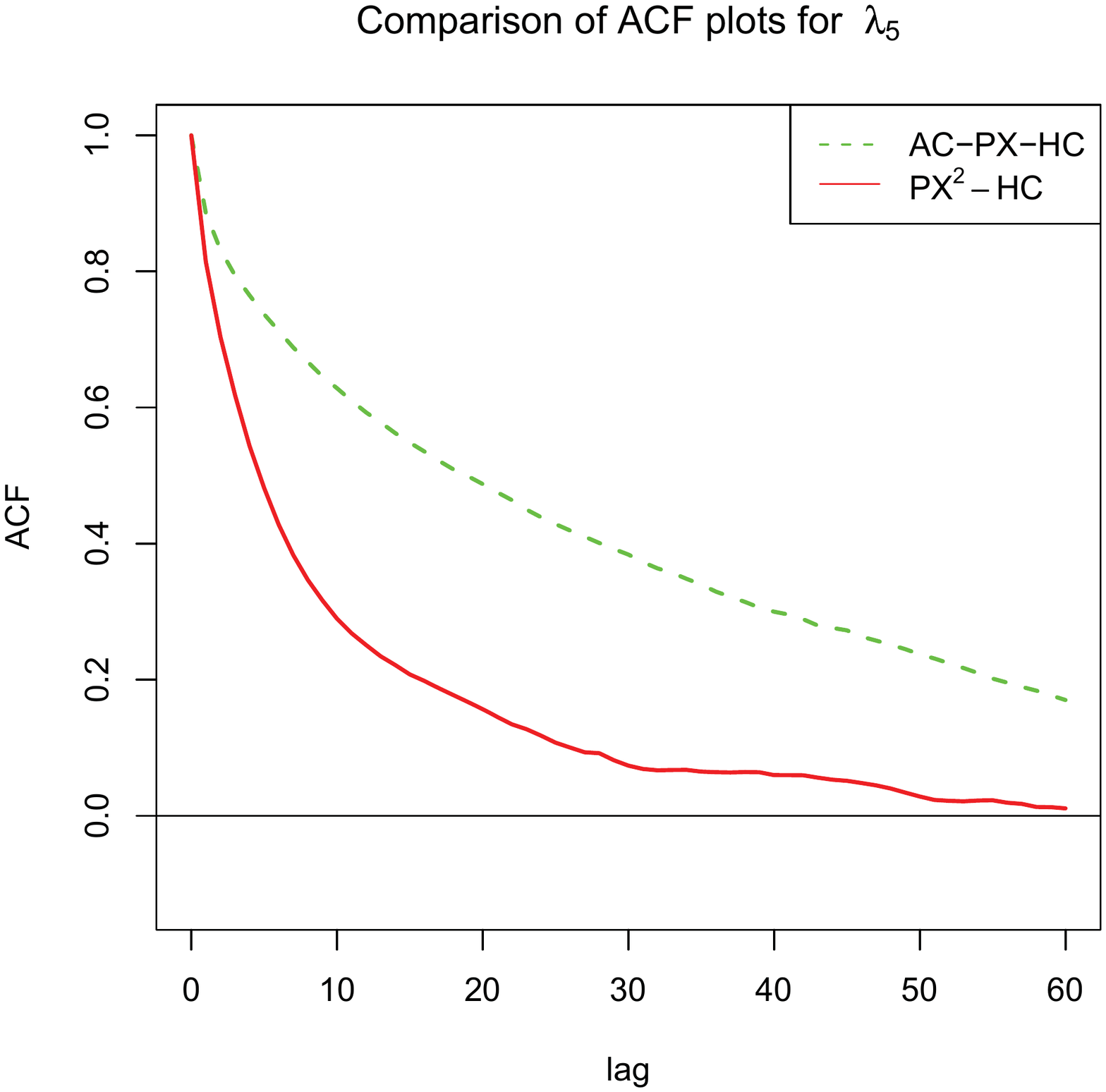}

\label{binary_scheme_acf_plot}
\end{figure}

\begin{figure}
  \caption{{\em   Autocorrelation functions (ACF) for the  standard Gibbs (SG) sampling scheme (left panel) and the proposed $PX^2$-HC sampling scheme (right panel),  averaged over 100 simulation replicates.    
 The autocorrelation is based on the posterior draws for the three factor loadings $\lambda_j$s ($j=1, 2, 3$) for the three continuous  phenotypes, as discussed in Section 5.1 and described in Table 1. }}
 \bigskip
 \bigskip
 
\fbox{\includegraphics[trim=0cm 5cm 0cm 3cm, clip=true,height=3.5in,width=0.45\textwidth]{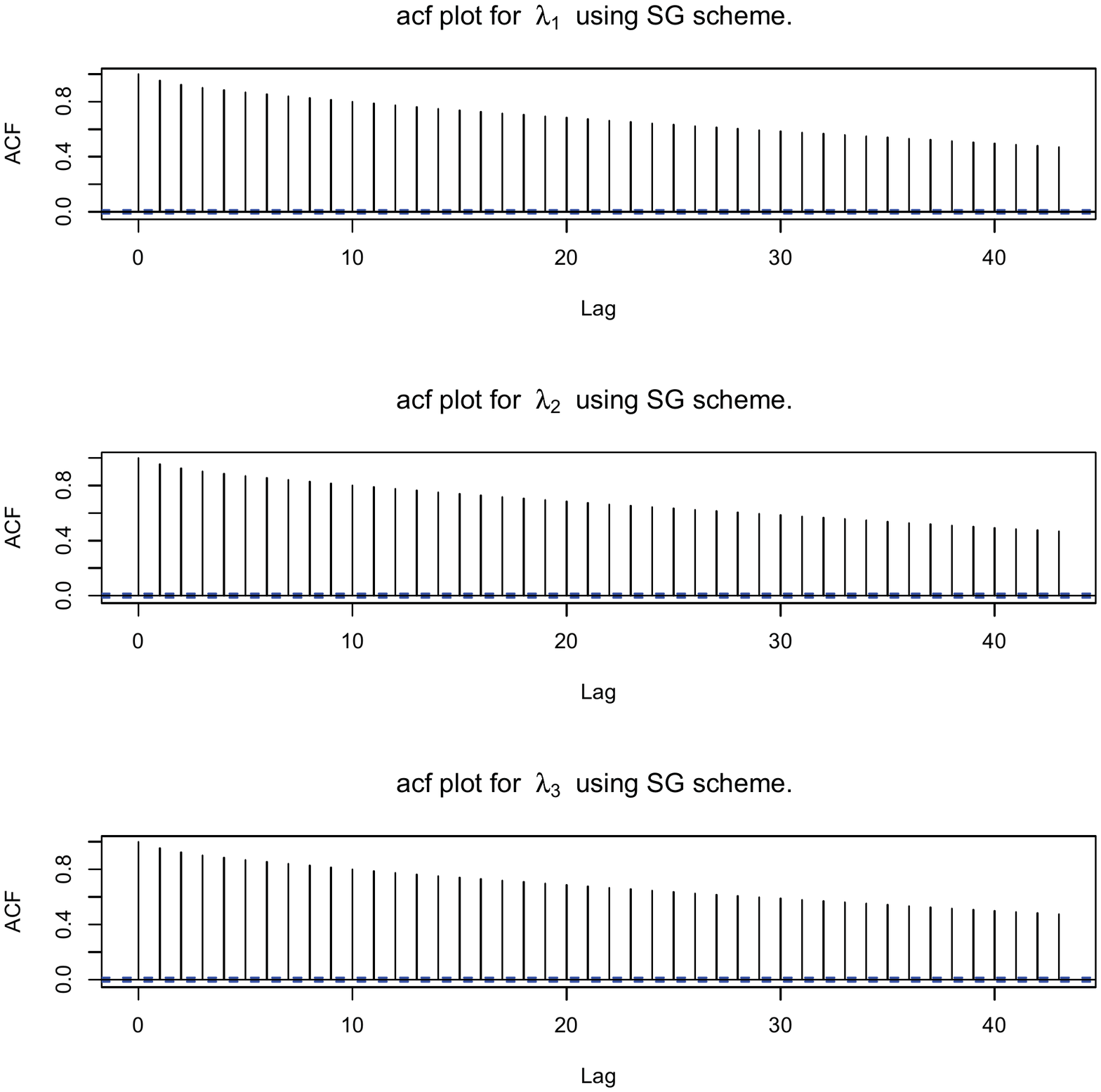}}
\fbox{\includegraphics[trim=0cm 5cm 0cm 3cm, clip=true,height=3.5in,width=0.45\textwidth]{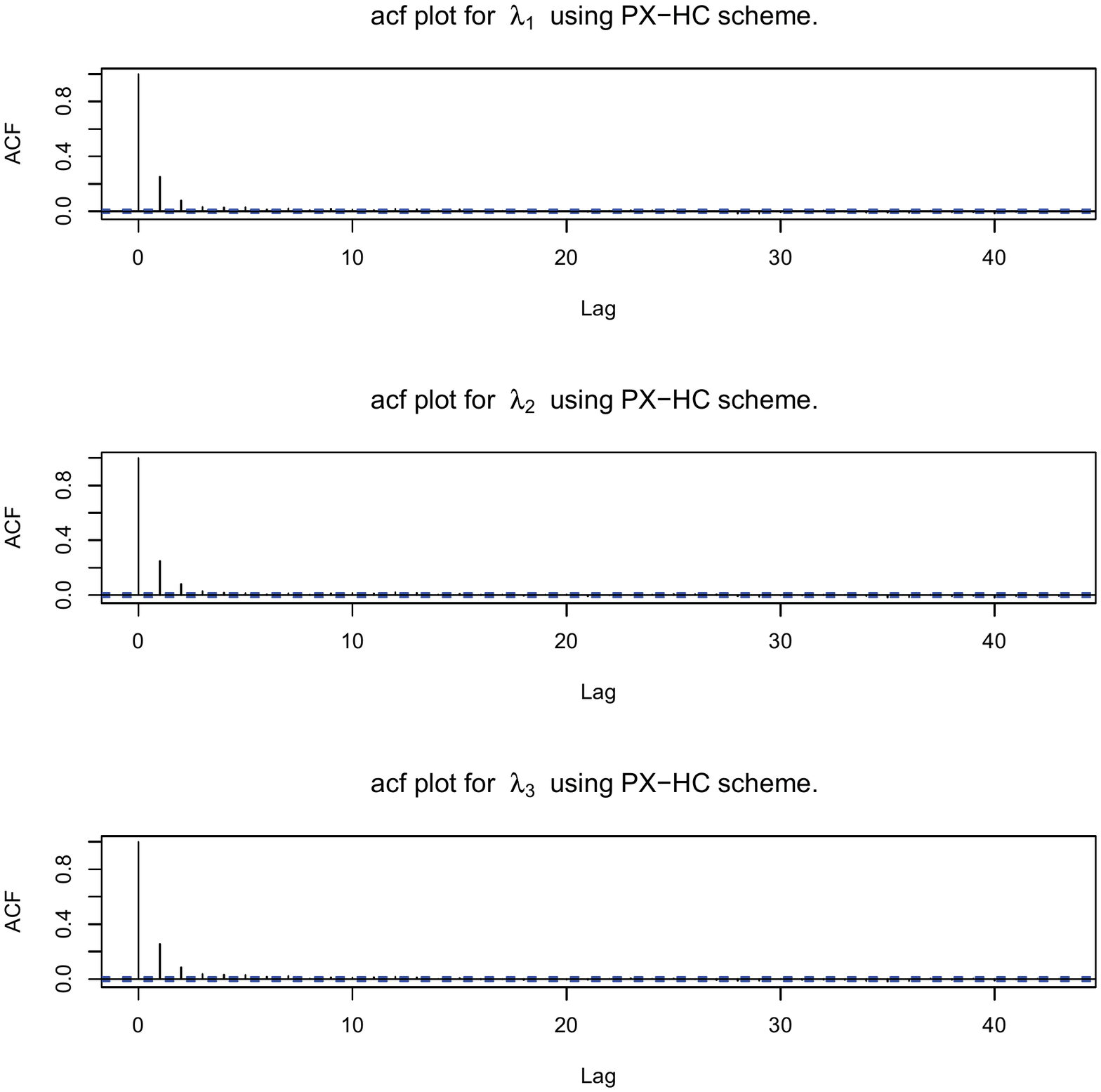}}
\label{ACF compare 2}
\end{figure}

\begin{landscape}

\begin{figure}
\includegraphics[trim=0cm 5cm 0cm 5cm,height=0.4\textwidth,width=0.25\textwidth]{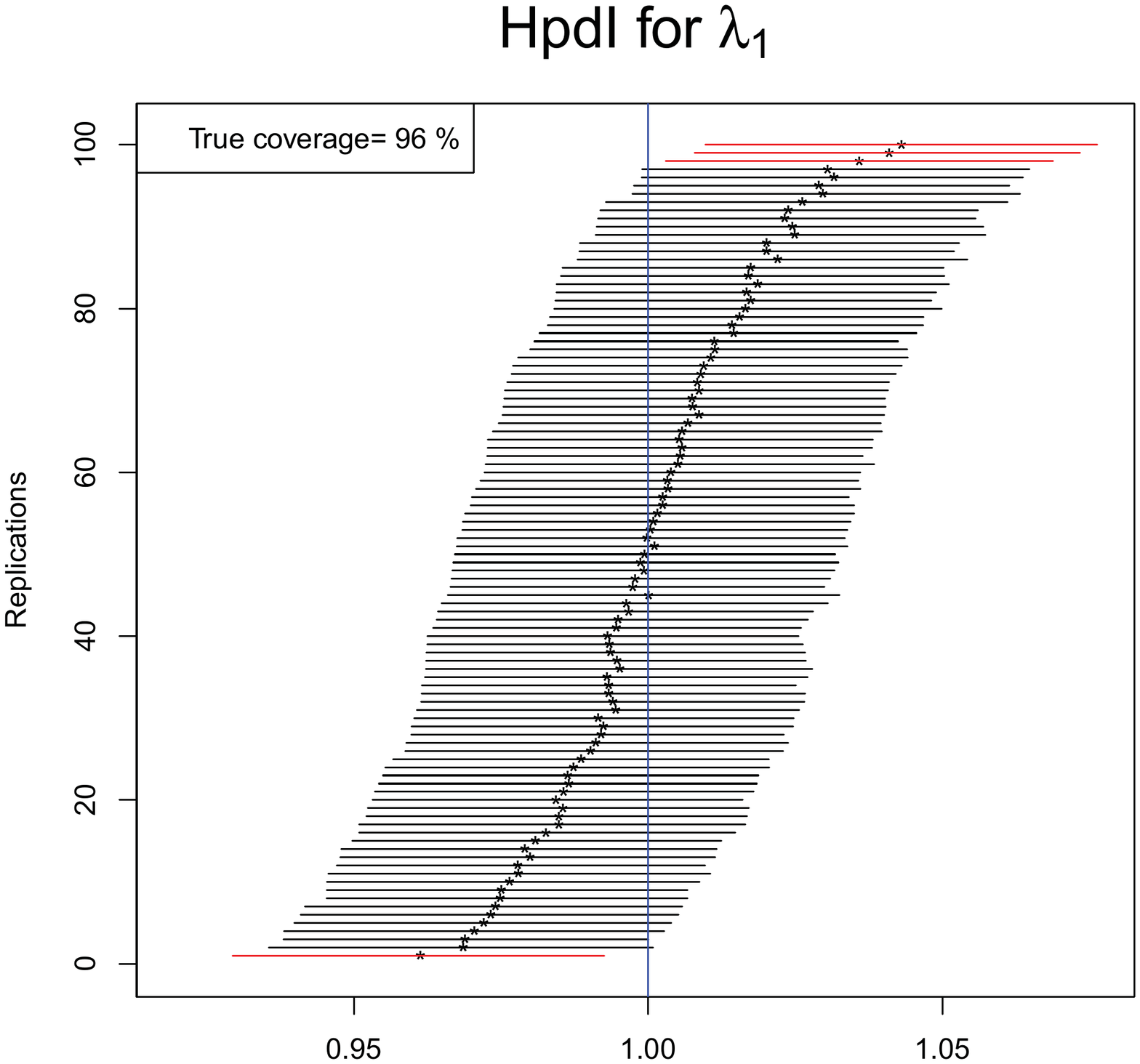}
\includegraphics[trim=0cm 5cm 0cm 5cm,height=0.4\textwidth,width=0.25\textwidth]{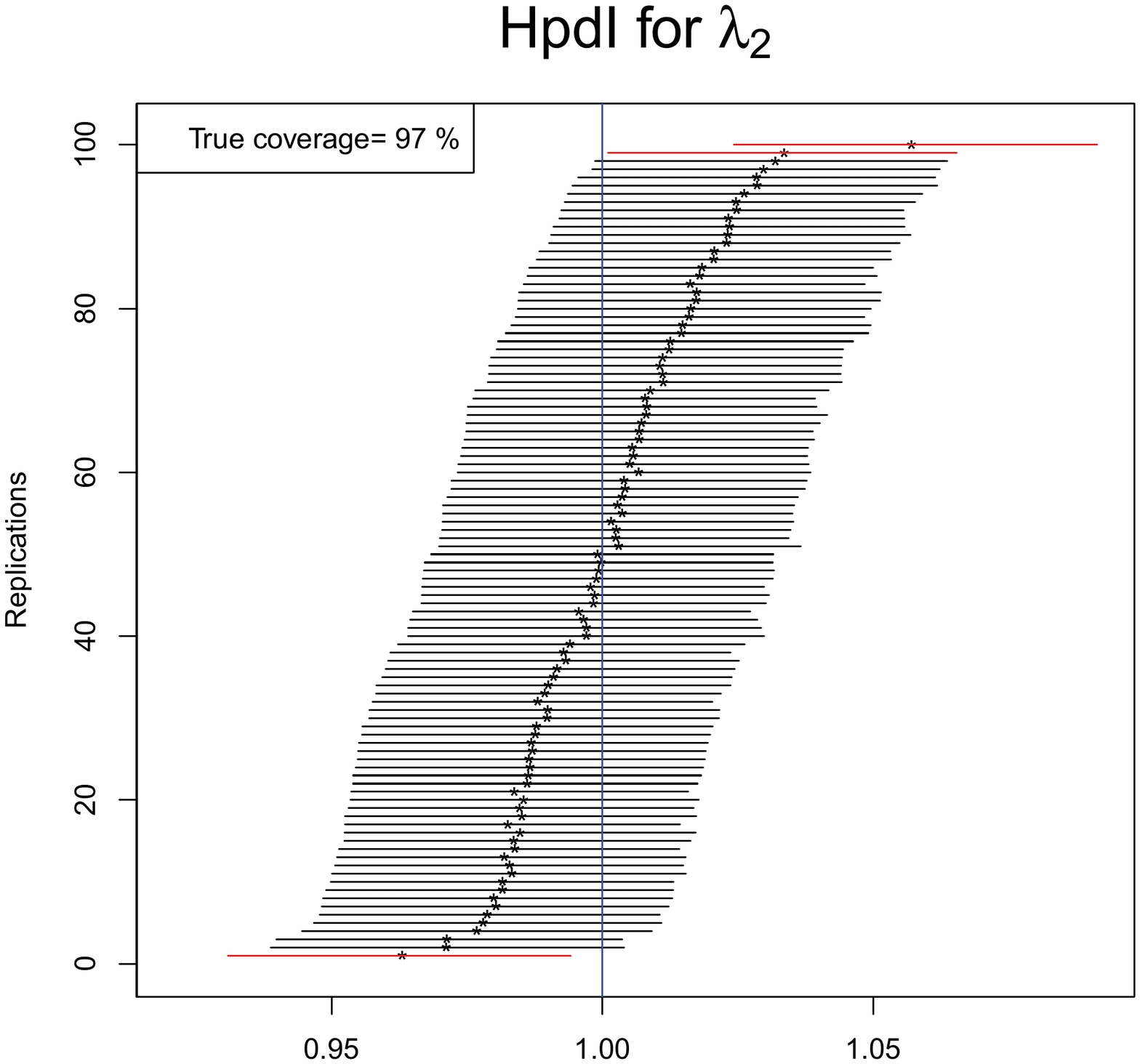}
\includegraphics[trim=0cm 5cm 0cm 5cm,height=0.4\textwidth,width=0.25\textwidth]{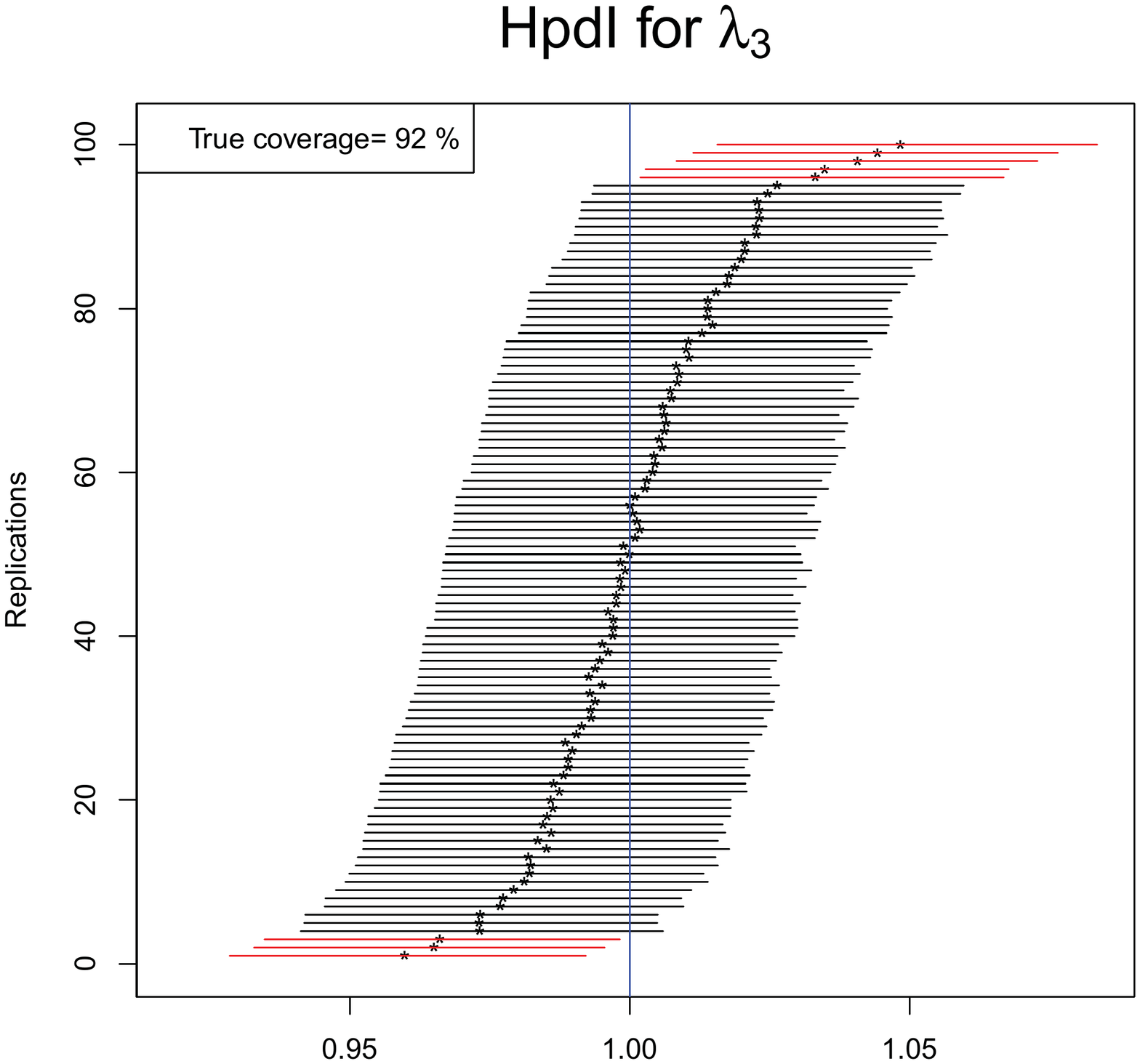}
\includegraphics[trim=0cm 5cm 0cm 5cm,height=0.4\textwidth,width=0.25\textwidth]{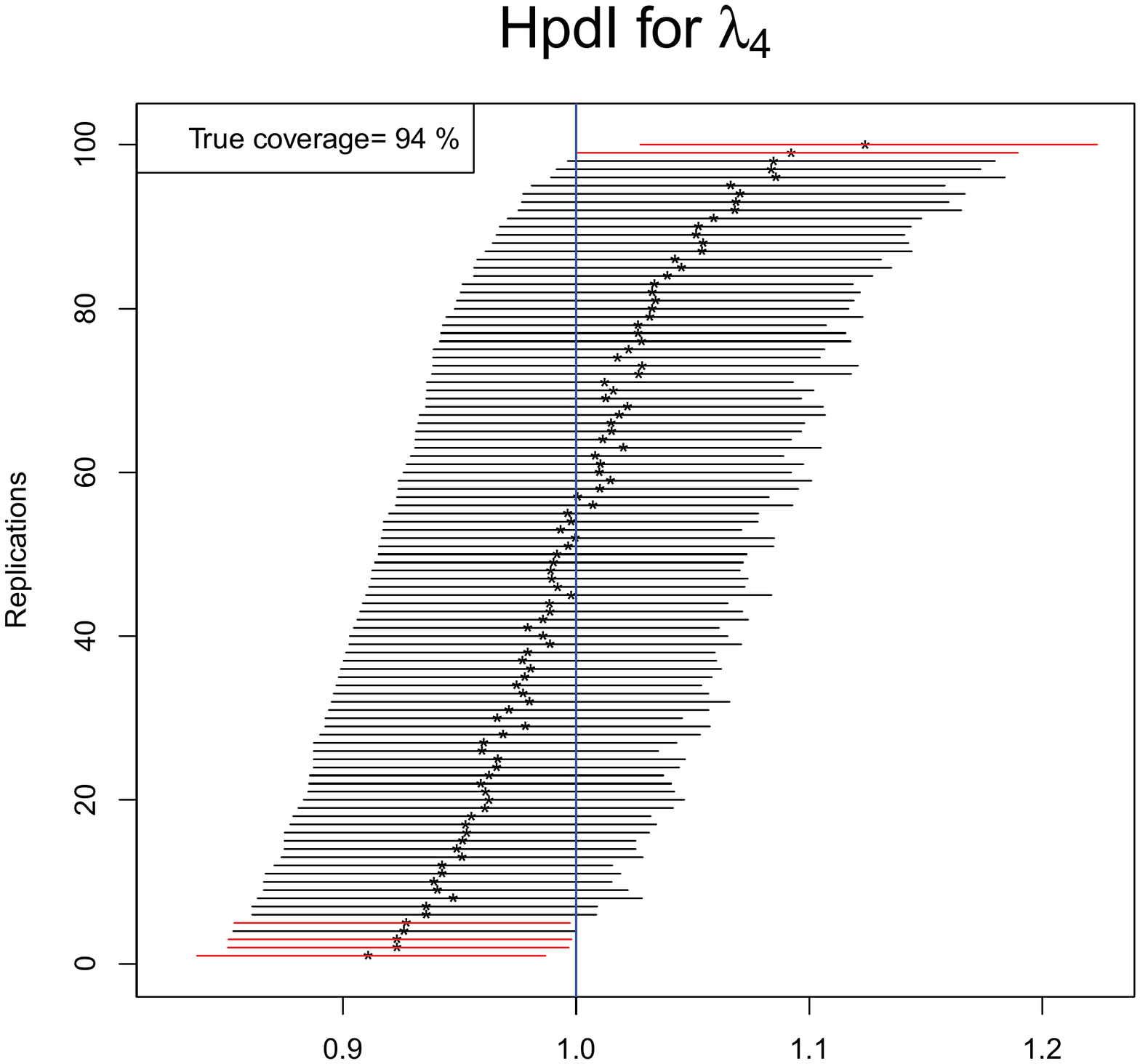}
\includegraphics[trim=0cm 5cm 0cm 5cm,height=0.4\textwidth,width=0.25\textwidth]{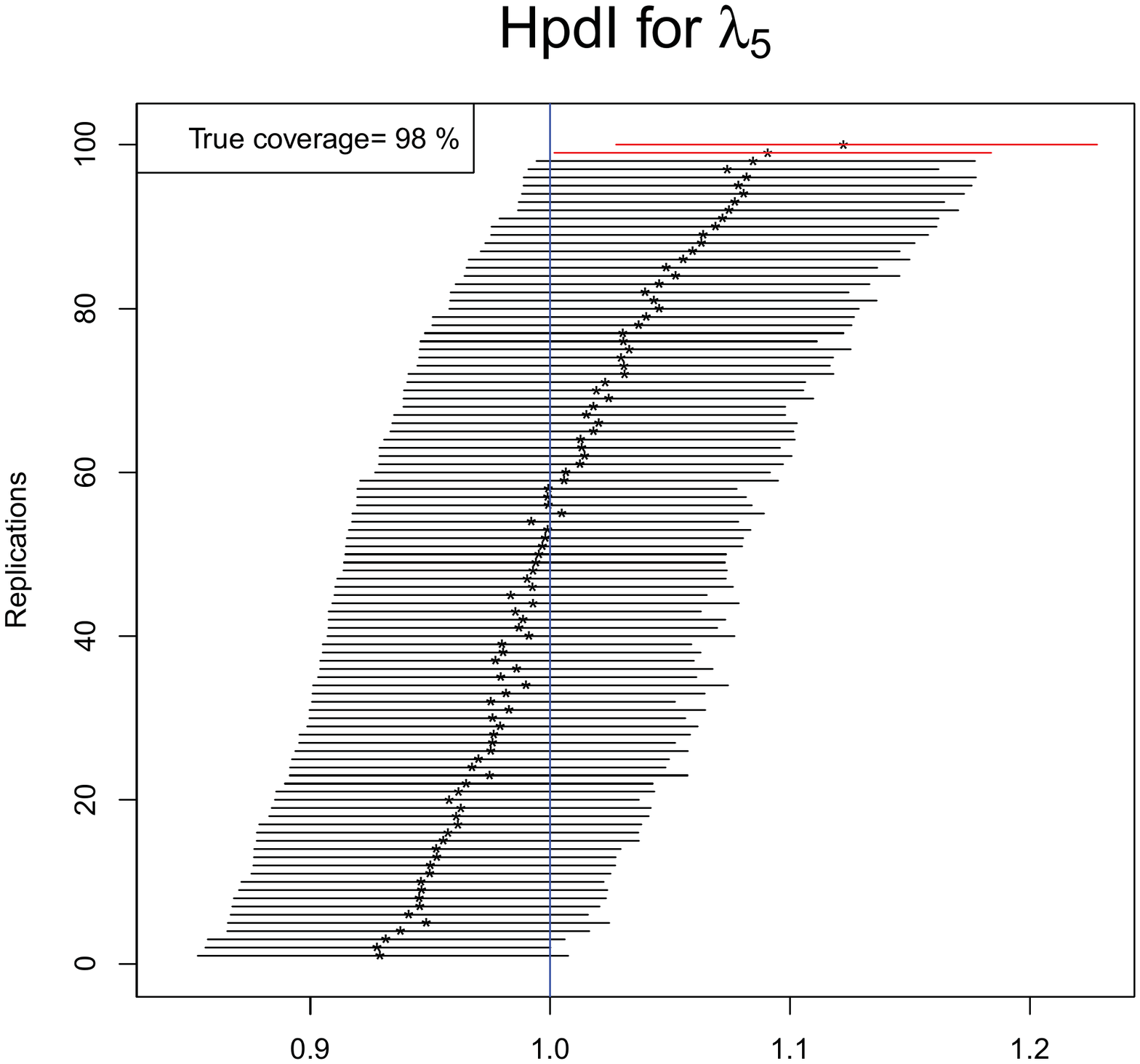}
 \caption{{\em The 95\% highest posterior density intervals (HpdI) for each of the five factor loadings $\lambda_j$s.
 The HpdI from the 100 simulation replicates are ordered by the lower bounds of the intervals. 
   The vertical line is
the true value and ``X"  marks the estimate from each simulation replicate.
The red HpdI do not cover the true value.
 The empirical coverages of the HpdI are 94\%, 95\%, 91\%, 94\%, 95\% (from left to right), respectively, for the five phenotypes $y_1$, $y_2$, $y_3$, $y_4$ and $y_5$ as simulated in Section 5.1 and described in Table 1. The legend indicates the empirical coverage  estimated from the 100 intervals. }}
\label{HPDI fig}
\end{figure}
\end{landscape}

 \end{document}